\documentclass[preprintnumbers,amsmath,amssymb,pre,floatfix,showkeys,superscriptaddress,nofootinbib,secnumarabic]{revtex4}

\usepackage[english]{babel}
\usepackage[latin2]{inputenc}
\usepackage[T1]{fontenc}
\usepackage{color}
\usepackage{amsmath}
\usepackage{amssymb}
\usepackage{ae,aecompl}
\usepackage{enumerate}
\usepackage{epsfig}
\usepackage{graphics}
\usepackage{graphicx}
\usepackage{pifont}
\usepackage[sort&compress]{natbib}
\usepackage{wasysym}
\usepackage{rotating}
\usepackage{multirow}

\newcommand{\addtallfig}[4]
{
\begin{figure*}[!th]
\centerline{\includegraphics[height=240pt,angle=-90,trim=130 0 -50 320]{#2#3}}
\caption{#4}
\label{fig:#1}
\end{figure*}
}

\newcommand{\addsmallfig}[4]
{
\begin{figure}[!ht]
\centerline{\includegraphics[width=155pt,angle=-90]{#2#3}}
\caption{#4}
\label{fig:#1}
\end{figure}
}

\newcommand{\addwidefig}[4]
{
\begin{figure*}[!tp]
\centerline{\includegraphics[width=170pt,angle=-90,trim=180 0 50 30]{#2#3}}
\caption{#4}
\label{fig:#1}
\end{figure*}
}

\newcommand{\addtwofigs}[5]
{
\begin{figure*}[!t]
\centerline{\includegraphics[height=240pt,angle=-90,trim=130 0 -50 320]{#2#3}\includegraphics[height=240pt,angle=-90,trim=130 0 -50 320]{#2#4}}
\caption{#5}
\label{fig:#1}
\end{figure*}
}

\newcommand{\paddtwofigs}[5]
{
\begin{figure*}[!p]
\centerline{\includegraphics[height=240pt,angle=-90,trim=130 0 -50 320]{#2#3}\includegraphics[height=240pt,angle=-90,trim=130 0 -50 320]{#2#4}}
\caption{#5}
\label{fig:#1}
\end{figure*}
}

\catcode`\_=12\catcode`\_=8

\def\ev #1{\left\langle #1 \right\rangle}

\def\MO{\mathrm{MO}}
\def\CA{\mathrm{CA}}
\def\LO{\mathrm{LO}}
\def\MOO{\mathrm{MO}^0}
\def\CAO{\mathrm{CA}^0}
\def\LOO{\mathrm{LO}^0}
\def\DeltaR{\Delta^\mathrm{R}}
\def\etal{{\it et al}. }
\def\ccc#1;#2{\left\langle #1 \left\vert #2 \right.\right\rangle}
\def\ev #1{\left\langle #1 \right\rangle}

\begin{document}

\preprint{}
\title{The price impact of order book events: market orders, limit orders and cancellations}
\author{Zolt\'an Eisler}
\affiliation{Capital Fund Management, Paris, France}
\author{Jean-Philippe Bouchaud}
\affiliation{Capital Fund Management, Paris, France}
\author{Julien Kockelkoren}
\affiliation{Capital Fund Management, Paris, France}

\date{\today}

\begin{abstract}
While the long-ranged correlation of market orders and their impact on prices has been relatively well studied in the literature, the
corresponding studies of limit orders and cancellations are scarce. We provide here an empirical study of the cross-correlation between
all these different events, and their respective impact on future price changes. We define and extract from the data the ``bare''
impact these events would have, if they were to happen in isolation. For large tick stocks, we show that a model where the bare impact 
of all events is permanent and non-fluctuating is in good agreement with the data. For small tick stocks, however, bare impacts must contain a
history dependent part, reflecting the internal fluctuations of the order book. We show that this effect can be accurately described 
by an autoregressive model on the past order flow. This framework allows us to decompose the impact of an event into three parts: 
an instantaneous jump component, the modification of the future rates of the different events, and
the modification of the jump sizes of future events. We compare in detail the present formalism with the temporary impact model
that was proposed earlier to describe the impact of market orders when other types of events are not observed. Finally, we extend the model to
describe the dynamics of the bid-ask spread.
\end{abstract}


\maketitle

\tableofcontents

\vfill\eject

\section{Introduction}

The relation between order flow and price changes has attracted considerable attention in the recent years \cite{hasbrouck.book, mike.empirical, bouchaud.subtle, bouchaud.molasses, lyons.book, bouchaud.review}. To the investors' dismay, trades on average impact the price in the direction of their transactions, i.e. buys push the price up and sells drive the price down. Although this
sounds very intuitive, a little reflection shows that such a statement is far from trivial, for any buy trade in fact meets a sell trade, 
and vice-versa! On the other hand, there must indeed be a mechanism allowing information to be included into and reflected by prices. This is well illustrated by the Kyle model \cite{kyle.model}, where the trading of an insider progressively reveals his information by impacting the price. Traditionally, the above ``one sell for one buy'' paradox is resolved by arguing that there are in fact two types of traders coexisting in the ecology of financial markets: (i) ``informed'' traders who place market orders for
immediate execution, at the cost of paying half the bid-ask spread, and (ii) uninformed (or less informed) market makers who provide liquidity by placing limit orders on both sides of the order book, hoping to earn part of the bid-ask spread. In this setting, there is indeed an asymmetry between a buyer, 
placing a market order at the ask, and the corresponding seller with a limit order at the ask, and one can speak about a well defined impact of buy/sell (market) orders. The impact of market orders has therefore been empirically studied in great detail since the early nineties. As reviewed below, many surprising results have been obtained, such
as a very weak dependence of impact on the volume of the market order, the long-range nature of the sign of the trades, and the resulting non-permanent, power-law decay of impact.

The conceptual problem is that the distinction between informed trader and market maker is no longer obvious in the present electronic markets, where each participant can place both limit and market orders, depending on his own strategies, the current state of the order book, etc. Although there is still an asymmetry between a buy market order and a sell limit order that enables one to define the direction of the trade, ``informed'' traders too may choose to place limit orders, aiming to decrease execution costs. Limit orders must therefore also have an impact: adding a buy limit order induces extra upwards pressure, and cancelling a buy limit order decreases this pressure. Surprisingly, there are very few quantitative studies of the impact of these orders -- partly due
to the fact that more detailed data, beyond trades and quotes, is often needed to conduct such studies. As this paper was under review, we became aware of ref. \cite{hautsch.limit}, where a similar
empirical study of the impact of limit orders is undertaken.

The aim of the present paper is to provide a unified framework for the description of the impact of {\it all} order book events, at least at the best limits: market orders, limit orders and cancellations. We study the correlations between all events types and signs. Assuming an additive model of impact, we map out from empirical data (consisting purely of trades and quotes information) the average individual impact of these orders. We find that the impact of limit orders is similar (albeit somewhat smaller) to that of market orders. 

We then compare these results to a simple model which assumes that all impacts are permanent in time. This works well for large tick stocks, 
for which the bid-ask spread is nearly constant, with no gaps in the order book. The discrepancies between this simple model and data from small tick stocks are then scrutinized in detail and attributed to the {\it history dependence} of the impact, which we are able to model successfully using a linear regression of the gaps on the past order flow. Our final model is specified in Sec. \ref{sec:gap}, Eq. \eqref{eq:finalmodel}. This framework allows us to measure more accurately the average impact of all types of orders, and to assess precisely the importance of impact fluctuations due to changes in the gaps behind the best quotes.

We want to insist on the fact that our study is mostly empirical and phenomenological, in the sense that we aim at establishing some stylized facts and building a parsimonious mathematical model to describe them without at this stage referring to any precise economic reasoning about the nature and motivations of the agents who place the orders. For recent papers along this latter direction, see e.g. \cite{henderstott.algo, biais.weill}. We, however, tend not to believe in the possibility, for now, to come up with a model that economists would like, with 
agents, equilibrium, etc. It seems to us that before achieving this, a more down to earth (but comprehensive) description of the data is needed, on which intuition can be built. This is what we try to provide in this paper. We are also aware of the cultural gap between communities, and that our work will appear to some researchers as ``eyeball econometrics''. We, however, are deeply convinced that a graphical representation of data is needed to foster intuition, before any rigourous calibration is attempted.

The outline of this paper is as follows. We first review (Sec. \ref{sec:review}) the relevant results on the impact of market orders and set the mathematical framework within which we will analyze our order book event data. We explain in particular why the market order impact function measured in previous studies is in fact ``dressed'' by the impact of other events (limit orders, cancellations), and by the history dependence of the impact. We also relate our formalism to Hasbrouck's Vector Autoregression framework. We then turn to the presentation of the data we have analyzed (Sec. \ref{sec:data}), and of the various correlation functions that one can measure (Sec. \ref{sec:cr}). From these we determine the individual (or ``bare''), lag-dependent impact functions of the different events occuring at the bid price or at the ask price (Sec. \ref{sec:G}). We introduce a simplified model where these impact functions are constant in time, and show that this gives an good approximate account of our data for large tick stocks, while significant discrepancies appear for small tick stocks (Sec. \ref{sec:perm}). The systematic differences are explained by the dynamics of order flow deeper in the book, which can
be modeled as a history dependent correction to the linear impact model (Sec. \ref{sec:gap}, see Eq. \eqref{eq:finalmodel}). Our results are summarized in the conclusion, with open issues that would deserve more detailed investigation. In the Appendices we also show how the bid-ask spread dynamics can be accounted for within the framework introduced in the main text (Appendix \ref{app:spread}) and some supplementary information concerning the different empirical correlations that can be measured (Appendix \ref{app:data}).

\section{Impact of market orders: a short review}
\label{sec:review}

\subsection{The transient impact model}

Quantitative studies of the price impact of market orders have by now firmly established a number of stylized facts, some of which appear rather
surprising at first sight. The salient points are (for a recent review and references, see \cite{bouchaud.review}):
\begin{itemize}
\item Buy (sell) trades on average impact the price up (down). In other words, there is a strong correlation between price returns over a given 
time interval and the market order imbalance on the same interval.
\item The impact curve as a function of the volume of the trade is strongly concave. In other words, large volumes impact the price only marginally
more than small volumes.
\item The sign of market orders is strongly autocorrelated in time. Despite this, the dynamics of the midpoint is very close to being purely diffusive.
\end{itemize}
A simple model encapsulating these empirical facts assumes that the mid-point price $p_t$ can be written at (trade) time $t$ as a linear superposition
of the impact of past trades \cite{bouchaud.subtle, bouchaud.molasses}:\footnote{In the following, we only focus on price changes over small
periods of time, so that the following additive model is adequate. For longer time scales, one should worry about multiplicative effects, which
in this formalism would naturally arise from the fact that the bid-ask spread, and the gaps in the order book, are a fraction of the price. Therefore,
the impact itself, ${\cal G}$, is expected to be proportional to a moving average of the price. See \cite{bouchaud-potters.book} for a discussion of this point.}
\begin{equation}
    p_t = \sum_{t'<t}\left[{G}(t-t')\epsilon_{t'}v_{t'}^\theta + n_{t'} \right] + p_{-\infty},
    \label{eq:ptMO}
\end{equation}
where $v_{t'}$ is the volume of the trade at time $t'$, $\epsilon_{t'}$ the sign of that trade ($+$ for a buy, $-$ for a sell), and $n_t$ is an
independent noise term that models any price change not induced by trades (e.g. jumps due to news). The exponent $\theta$
is small; the dependence in $v$ might in fact be logarithmic ($\theta \to 0$). The most important object in the above equation is the function ${G}(t-t')$ 
which describes the temporal evolution of the impact of a single trade, which can be called a `propagator': how does the impact of the trade at 
time $t'<t$ propagate, on average, up to time $t$? We discuss in section 2.4 below how Eq. (\ref{eq:ptMO}) is related to Hasbrouck's VAR model \cite{hasbrouck.var, hasbrouck.book}

An important result, derived in \cite{bouchaud.subtle}, is that ${G}(t-t')$ must decay with time in a very specific way, 
such as to off-set the autocorrelation of the trades, and maintain the (statistical) efficiency of prices. Clearly, if ${G}(t-t')$ did not decay at all, the returns would simply be proportional to the sign of the trades, and therefore would themselves be strongly autocorrelated in time. The resulting price dynamics would then be
highly predictable, which is not the case. Conversely, if ${G}(t-t')$ decayed to zero immediately, the price as given by Eq. \eqref{eq:ptMO} 
would oscillate within a limited range, and the long-term volatility would be zero. The result of \cite{bouchaud.subtle} is that if the correlation of signs $C(\ell) = \langle \epsilon_t \epsilon_{t+\ell} \rangle$ decays at large $\ell$ as $\ell^{-\gamma}$ with $\gamma < 1$ (as found empirically), then ${G}(t-t')$ must decay 
as $|t-t'|^{-\beta}$ with $\beta = (1-\gamma)/2$ for the price to be exactly diffusive at long times. The impact of single trades is therefore predicted to decay as a power-law (at least up to a certain time scale), at variance with simple models that assume that the impact decays exponentially to a non-zero ``permanent'' value. More generally, one can use the empirically observable impact function ${\cal R}(\ell)$, defined as:
\begin{equation}
    {\cal R}(\ell) = \langle (p_{t+\ell}-p_t) \cdot \xi_t \rangle,
\end{equation}
and the time correlation function $C(\ell)$ of the variable $\xi_t=\epsilon_{t}v_{t}^\theta$ to map out, numerically, the complete shape of ${G}(t-t')$. This was done 
in \cite{bouchaud.molasses}, using the exact relation: 
\begin{equation}
\label{eq:RCG}
    {\cal R}(\ell)  = \sum_{0 < n\leq \ell} {G}(n) C(\ell-n) + \sum_{n>\ell} {G}(n) C(n-\ell) - \sum_{n>0} {G}(n) C(n).
\end{equation}
This analysis is repeated in a more general setting below (see Sec. \ref{sec:G} and Eq. \eqref{eq:pt}). The above model, however, is approximate 
and incomplete in two, interrelated ways. 
\begin{itemize}
\item First, Eq. \eqref{eq:ptMO} neglects the fluctuations of the impact: one expects that ${\cal G}(t' \to t)$, which is the impact of trade at some time $t'$ measured until a later time $t$, to depend both on $t$ and $t'$ and not only on $t-t'$. Its formal definition is given by:
\begin{equation}
  {\cal G}(t' \to t) = \frac{\partial p_t}{\partial \xi_{t'}}, \qquad \xi_t \equiv \epsilon_{t}v_{t}^\theta.  
    \label{eq:Gdef}
\end{equation}
Impact can indeed be quite different depending on the state of the order book and the market conditions at $t'$. As a consequence, if one blindly uses 
Eq. \eqref{eq:ptMO} to compute the second moment of the price difference, $D(\ell)=\langle (p_{t+\ell}-p_t)^2 \rangle$, with a non-fluctuating ${G}(\ell)$ calibrated to 
reproduce the impact function ${\cal R}(\ell)$, the result clearly underestimates the empirical price variance: see Fig. \ref{fig:vol_test_fair_just_MO_smalltick}.
Adding a diffusive noise $n_t \neq 0$ would only shift $D(\ell)/\ell$ upwards, but this is insufficient to reproduce the empirical data. 
\item Second, other events of the order book can also change the mid-price, such as limit orders placed inside the bid-ask spread, or cancellations of all the volume at the bid or the ask. These events do indeed contribute to the price volatility and should be explicitly included in the description. A simplified description of price changes  in terms of market orders only attempts to describe other events of the order book in an effective way, through the non-trivial time dependence of ${G}(\ell)$. 
\end{itemize}

\addsmallfig{vol_test_fair_just_MO_smalltick}{}{vol_test_fair_just_MO_smalltick}{$D(\ell)/\ell$ and its approximation with the temporary impact model with only trades as events, with $n_t=0$ and for small tick stocks. Results are shown when assuming that all trades have the same, non fluctuating, impact $G(\ell)$, calibrated to reproduce ${\cal R}(\ell)$. This simple model accounts for $\sim 2/3$ of the long term volatility. Other events and/or the fluctuations of impact ${G}(\ell)$ must therefore contribute to the market volatility as well.}

\subsection{History dependence of the impact function}

Let us make the above statements more transparent on toy-models. First, the assumption of a stationary impact function ${\cal G}(t'\to t)={G}(t-t')$ is clearly an approximation. The past order flow ($< t'$) should affect the way the trade at time $t'$ impacts the price, or, as argued by Lillo and
Farmer, that liquidity may be history dependent \cite{lillo-farmer.qf,gerig.phd, bouchaud.review}. Suppose for simplicity that the variable $\xi_t=\epsilon_{t}v_{t}^\theta$ is Gaussian (which turns out to be a good approximation) and that its impact is permanent but history dependent. If we assume that the past order flow has a small influence on the impact, we can formally expand ${\cal G}$ in powers of all past $\xi$'s to get:
\begin{equation}
    {\cal G}(t' \to t) = G_0 + G_1 \sum_{t_1 < t'} g_1(t'-t_1) \xi_{t_1} + G_2  \sum_{t_1,t_2 < t'} g_2(t'-t_1;t'-t_2) \xi_{t_1}\xi_{t_2} + \cdots .
\end{equation}
If buys and sells play a symmetric role, $G_1 = 0$. 
Using the fact the $\xi$'s are Gaussian with zero mean, one finds that the impact function ${\cal R}(\ell)$ within this toy-model is given by:
\begin{equation}
    {\cal R}(\ell) = \sum_{0 < n\leq \ell} C(n) \left[G_0 + G_2 \sum_{n_1,n_2 > 0} g_2(n_1;n_2) C(n_1-n_2)\right] + 
    2 G_2 \sum_{0 < n\leq \ell} \sum_{n_1,n_2 > 0} g_2(n_1;n_2) C(n_1) C(n-n_2). 
\end{equation}
If one compares this expression with Eq. \eqref{eq:RCG} to extract an effective propagator ${G}(\ell)$, it is clear that the resulting solution will 
have some non-trivial time dependence induced by the third term, proportional to $G_2$.

\subsection{The role of hidden events}

Imagine now that two types of events are important for the dynamics
of the price. Events of the first type are characterized by a random variable $\xi_t$ (e.g., $\xi_t=\epsilon_{t}v_{t}^\theta$ in the above
example), whereas events of the second type (say limit orders) are charaterized by another random variable $\eta_t$. The ``full'' dynamical equation for the price is given
by:
\begin{equation}
    p_t = \sum_{t'<t} G_1(t-t') \xi_{t'} + \sum_{t'<t} G_2(t-t') \eta_{t'} + p_{-\infty},
    \label{eq:pttoy1}
\end{equation}
Imagine, however, that events of the second type are {\it not} observed. If for simplicity $\xi$ and $\eta$'s are correlated 
Gaussian random variables, one can always express the $\eta$'s as linear superposition of past $\xi$'s and find a model in terms of $\xi$'s only, plus an uncorrelated `noise' component $n_t$ coming from the unobserved events:
\begin{equation}
    p_t = \sum_{t'<t} \left[G_1(t-t') \xi_{t'} + G_2(t-t') \sum_{t'' \leq t'} \Xi(t'-t'') \xi_{t''}\right] + n_t + p_{-\infty}.
    \label{eq:pttoy2}
\end{equation}
$\Xi$ is the linear filter allowing to predict the $\eta$'s in terms of the past $\xi$'s. It can be expressed in a standard way in terms of the correlation function of the $\eta$'s and the cross-correlation between $\xi$'s and $\eta$'s. Notice that the previous equation can be recast in the form
of Eq. (\ref{eq:ptMO}) plus noise, with an effective propagator ``dressed'' by the influence of the unobserved events:
\begin{equation}
    {G}(\ell) = G_1(\ell)  +  \sum_{0 < \ell' \leq \ell} G_2(\ell') \Xi(\ell-\ell').
    \label{eq:pttoy3}
\end{equation}
From this equation, it is clear that a non-trivial dependence of ${G}$ can arise even if the `true' propagators $G_1$ and $G_2$ are time 
independent -- in other words the decay of the impact of a single market order is in fact a consequence of the interplay of market and limit 
order flow. As a trivial example, suppose both bare propagators are equal and constant in time ($G_1(\ell)=G_2(\ell)=G$)
and $\eta_t \equiv -\xi_t$, $\forall t$. This means that the two types of events impact the price but exactly cancel
each other. Then, $\Xi(\ell)=\delta_{\ell,0}$ and ${G}(\ell) \equiv 0$, as it should: the dressed impact of events of the first type is zero.
This is an idealized version of the asymmetric liquidity model of Lillo and Farmer mentioned above \cite{lillo-farmer.qf}. 

The aim of this paper is to investigate a model of impact similar to Eqs. \eqref{eq:ptMO} and \eqref{eq:pttoy1}, but where a wider class of 
order book events are explicitly taken into account. This will allow us to extract the corresponding single event impact functions, and study
their time evolution. As a test for the accuracy of the model, the time behavior of other observables, such as the second moment
of the price difference should be correctly accounted for. We start by presenting the data and extra notations which will be useful in the sequel.
We then discuss the different correlation and response functions that can be measured on the data.

\subsection{Relation with Hasbrouck's VAR model}

At this stage, it is interesting to relate the above `propagator' framework encoded in Eq. (\ref{eq:ptMO}) and the econometric Vector Autoregressive (VAR) model proposed by Hasbrouck, and that
became a standard in the microstructure literature. 
In its original formulation, the VAR model is a joint linear regression of the present price return $r_t$ and signed volume $x_t=\epsilon_t v_t$ onto their past realisations, or more 
precisely:
\begin{eqnarray}
 \label{eq:var}
r_t = \sum_{t' < t} B_{rr}(t-t') r_{t'} + B_{xr}(0) x_t + \sum_{t' < t} B_{xr}(t-t') x_{t'} + n_r(t), \nonumber \\
x_t = \sum_{t' < t} B_{rx}(t-t') r_{t'} +  \sum_{t' < t} B_{xx}(t-t') x_{t'} + n_x(t),
\end{eqnarray}
where $n_{r,x}$ are i.i.d. noises and the $B(t-t')$ are regression coefficients, to be determined. Eq. (\ref{eq:ptMO}) can be seen as a special case of 
the VAR model, Eq. (\ref{eq:var}), provided the 
following identifications/modifications are made: a) $x_t \to \xi_t = \epsilon_t v_t^\theta$; b) the coefficients $B_{rr}(\ell)$ are assumed to be zero; 
c) since Eq. (\ref{eq:ptMO}) models prices and not returns, one has $G(\ell)=\sum_{0 \leq \ell' < \ell} B_{xr}(\ell')$. $G(\infty)$ is called 
the {\it information content} of a trade in Hasbrouck's framework; d) finally, although the autocorrelation $C(\ell)$ of the $\xi_t$ is measured, 
the dynamical model for the $\xi_t$ is left unspecified. 

Although the two models are very similar at the formal level, the major distinction lies in the interpretation, which in fact illustrates the difference between econometric models and ``microscopic'' models.
Whereas the VAR model postulates a general, noisy linear relation between two sets (or more) of variables and determines the coefficients via least squares,  we insist on a microscopic
mechanism that leads to an a priori structure of the model and an interpretation of the coefficients. Eq. (\ref{eq:ptMO}) is a causal model for impact, which postulates that the current price is a result of the impact of all past trades, plus some noise contribution $n_t$ that represents price moves not related to trades (for example, quote revisions after some news announcements). In this context, 
there is no natural interpretation for the $B_{rr}$ coefficients, which must be zero: past price changes cannot by themselves influence the present price, although these may of course affect the order flow $\xi_t$, which in turn impacts `physically' the price. On the other hand, the interpretation in terms of impact allows one to anticipate the limitations of the model and to suggest possible improvements, 
by including more events or by allowing for some history dependence, as discussed in the above subsections.

The aim of the present paper is to justify fully this modeling strategy by accounting for all events in the order book. In this case, the variation of the price can be tautologically 
decomposed in terms of these events, and the corresponding regression coefficients have a transparent interpretation. Furthermore, the limitations of a purely linear model appear very clearly as 
the history dependence of impact may induce explicit non-linearities (see Sec. \ref{sec:gap}).

\section{Data and notations}
\label{sec:data}

In this paper we analyze data on $14$ randomly selected liquid stocks traded at NASDAQ during the period 03/03/2008 -- 19/05/2008, a total of $53$ trading days (see Table \ref{tab:eventprobs} for details). The particular choice of market is not very important, many of our results were also verified on other markets (such as CME Futures, US Treasury Bonds and stocks traded at London Stock Exchange\footnote{The results for these markets are not
reproduced here, for lack of space, but the corresponding data is available on request.}), as well as on other time periods and they appear fairly robust.

We only consider the usual trading time between 9:30--16:00, all other periods are discarded. We will always use ticks ($0.01$ US dollars) as the units of price. We will use the name ``event" for any change that modifies the bid or ask price, or the volume quoted at these prices. Events deeper in the order book are unobserved and will not be described: 
although they do not have an immediate effect on the best quotes, our description will still be incomplete; in line with the previous section, we know that these unobserved events may ``dress'' the impact of the observed events. Furthermore, we note that the liquidity is fragmented and that the stocks we are dealing with are traded on multiple platforms. The activity on these other platforms will also
``dress'' the impact of observable events, in the sense of the previous section. This may account for some of the residual discrepancies reported below. 

Events will be used as the unit of time. This ``event time" is similar, but more detailed than the notion of transaction time used in many recent papers.
Since the dependence of impact on the volume of the trades is weak \cite{jones, bouchaud.review}, we have chosen to classify events not according to their volume but 
according to whether they change the mid-point or not. This strong dichotomy is another approximation to keep in mind. It leads to six possible types of events\footnote{{Our data also included a small number ($\approx 0.3 \%$) of marketable (or crossing) limit orders. In principle these could have been treated as a market order (and a consequent limit order for the remaining volume if there was any). Due to technical limitations we decided to instead remove these events and the related price changes.}}:
\begin{itemize}
\item market orders\footnote{To identify multiple trades that are initiated by the same market order, we consider as one market order all the trades in a given stock that occur on the same side of the book within a millisecond. Such a time resolution is sufficient for distinguishing trades initiated by different parties even at times of very intense trading activity.} that do not change the best price (noted $\MOO$) or that do (noted $\MO'$), 
\item limit orders at the current bid or ask ($\LOO$) or inside the bid-ask spread so that they change the price ($\LO'$), 
\item and cancellations at the bid or ask that do not remove all the volume quoted there ($\CAO$) or that do ($\CA'$). 
\end{itemize}
The upper index ' (``prime") will thus denote that the event changed any of the best prices, and the upper index $0$ that it did not. 
Abbreviations without the upper index ($\MO$, $\CA$, $\LO$) refer to both the price changing and the non-price changing event type. 
The type of the event occuring at time $t$ will be denoted by $\pi_t$.

Our sample of stocks can be divided into two groups: large tick and small tick stocks. Large tick stocks are such that the bid-ask spread is almost always equal to one tick, whereas small tick stocks have spreads that are typically a few ticks.
The behavior of the two groups is quite different, and this will be emphasized throughout the paper. For example, the events which change the best price have a relatively low probability for large tick stocks (about $3\%$ altogether), but not for small tick stocks (up to $40\%$). Table \ref{tab:eventprobs} shows a summary of stocks, and some basic statistics. 
Note that there is a number of stocks with intermediate tick sizes, which to some extent possess the characteristics of both groups. Technically, they can be treated in exactly the same way as small tick stocks, and all our results remain valid. However, for the clarity of presentation, we will not consider them explicitly in this paper.

Every event is given a sign $\epsilon_t$ according to its expected long-term effect on the price. For market orders this corresponds to usual order signs, i.e., $\epsilon_t = 1$ for buy market orders (at the ask price) and $-1$ for sell market orders (at the bid price). Cancelled sell limit orders and incoming buy limit orders both have $\epsilon_t = 1$, while others have $\epsilon_t=-1$. The above definitions are summarized in Table \ref{tab:eventtypes}. Note that the table also defines the gaps $\Delta_{\pi, \epsilon}$, which will be used later.

It will also be useful to define another sign variable corresponding to the ``side" of the event at time $t$, which will be denoted by $s_t$. It indicates whether the event $t$ took place at the bid ($s_t = -1$) or the ask ($s_t = 1$), thus:
\begin{equation}
s_t = \left\{
\begin{array}{rl}
\epsilon_t & \text{if } \pi_t = \MO^0 \mathrm{,\ } \MO' \mathrm{,\ } \CA^0 \mathrm{\ or\ } \CA'\\
-\epsilon_t & \text{if } \pi_t = \LO^0 \mathrm{\ or\ } \LO'
\end{array} \right.
\label{eq:sign}
\end{equation}

The difference between $\epsilon$ and $s$ is because limit orders correspond to the addition not the removal of volume, and thus they push prices away from the side of the book where they occur.

In the following calculations we will sometimes rely on indicator variables denoted as $I(\pi_{t}=\pi)$. This expression is $1$ if the event at $t$ is of type $\pi$ and zero otherwise. In other words, $I(\pi_{t}=\pi) = \delta_{\pi_{t}, \pi}$, where $\delta$ is the Kronecker-delta. We will also use the notation $\ev{\cdot}$ to denote the time average of the quantity between the 
brackets. For example, the unconditional probability of the event type $\pi$ can be, by definition, calculated as $P(\pi) = \ev{I(\pi_t = \pi)}$. 

The indicator notation, although sometimes heavy, simplifies the formal calculation of some conditional expectations. For example if a quantity $X_{\pi, t}$ depends on the event type $\pi$ and the time $t$, then its conditional expectation at times of $\pi$-type events is
$$\ev{X_{\pi_t, t}|\pi_t = \pi} = \frac{\ev{X_{\pi, t}I(\pi_t = \pi)}}{P(\pi)}.$$ 
Also, by definition 
\begin{equation}
\label{eq:trick2}
\sum_\pi I(\pi_t = \pi) = 1; \quad {\rm and} \quad \sum_\pi X_{\pi, t}I(\pi_t = \pi) = X_{\pi_t, t}.
\end{equation}

\begin{table*}[tbp]
    \begin{tabular}{|c|p{5cm}|p{3cm}|p{5.5cm}|}
	\hline
	$\pi$ & event definition & event sign definition & gap definition ($\Delta_{\pi, \epsilon}$) \\ \hline
	$\pi = \MOO$ & market order, volume $<$ outstanding volume at the best & $\epsilon = \pm 1$ for buy/sell market orders & $0$ \\ \hline
	$\pi = \MO'$ & market order, volume $\geq$ outstanding volume at the best & $\epsilon = \pm 1$ for buy/sell market orders & half of first gap behind the ask ($\epsilon=1$) or bid ($\epsilon=-1$) \\ \hline
	$\pi = \CAO$ & partial cancellation of the bid/ask queue & $\epsilon = \mp 1$ for buy/sell side cancellation & $0$ \\ \hline
	$\pi = \LOO$ & limit order at the current best bid/ask & $\epsilon = \pm 1$ for buy/sell limit orders & $0$ \\ \hline
	$\pi = \CA'$ & complete cancellation of the best bid/ask & $\epsilon = \mp 1$ for buy/sell side cancellation & half of first gap behind the ask ($\epsilon=1$) or bid ($\epsilon=-1$) \\ \hline
	$\pi = \LO'$ & limit order inside the spread & $\epsilon = \pm 1$ for buy/sell limit order & half distance of limit order from the earlier best quote on the same side \\ \hline
    \end{tabular}
    \caption{Summary of the $6$ possible event types, the corresponding definitions of the event signs and gaps.}
    \label{tab:eventtypes}
\end{table*}


\begin{table*}[tbp]
    \begin{tabular}{|c|c||c|c|c|c|c|c|c|c|c|}
	\hline
& \multirow{2}{*}{ticker} & \multirow{2}{*}{$P(\MOO)$} & \multirow{2}{*}{$P(\MO')$} & \multirow{2}{*}{$P(\CAO)$} & \multirow{2}{*}{$P(\LOO)$} & \multirow{2}{*}{$P(\CA')$} & \multirow{2}{*}{$P(\LO')$} & mean spread & mean price & time/event \\ 
& & & & & & & & (ticks) & (USD) & (sec) \\ \hline\hline
\multirow{7}{*}{\begin{sideways}large tick\end{sideways}} & AMAT & 0.042 & 0.011 & 0.39 & 0.54 & 0.0018 & 0.013 & 1.11 & 17.45 & 0.16 \\
& CMCSA & 0.040 & 0.0065 & 0.41 & 0.53 & 0.0021 & 0.0087 & 1.12 & 20.29 & 0.15 \\
& CSCO & 0.051 & 0.0085 & 0.40 & 0.53 & 0.0010 & 0.0096 & 1.08 & 67.77 & 0.10 \\
& DELL & 0.042 & 0.0087 & 0.40 & 0.54 & 0.0019 & 0.011 & 1.10 & 20.22 & 0.17 \\
& INTC & 0.052 & 0.0073 & 0.40 & 0.54 & 0.00080 & 0.0081 & 1.08 & 19.43 & 0.12 \\
& MSFT & 0.054 & 0.0087 & 0.40 & 0.53 & 0.0012 & 0.010 & 1.09 & 27.52 & 0.098 \\
& ORCL & 0.050 & 0.0090 & 0.40 & 0.54 & 0.0012 & 0.010 & 1.09 & 20.86 & 0.16 \\ \hline

\multirow{6}{*}{\begin{sideways}small tick\end{sideways}} & AAPL & 0.043 & 0.076 & 0.32 & 0.33 & 0.077 & 0.16 & 3.35 & 140.56 & 0.068 \\
& AMZN & 0.038 & 0.077 & 0.26 & 0.31 & 0.12 & 0.20 & 3.70 & 70.68 & 0.21 \\
& APOL & 0.042 & 0.080 & 0.24 & 0.33 & 0.11 & 0.20 & 3.78 & 55.24 & 0.40 \\
& COST & 0.054 & 0.069 & 0.27 & 0.36 & 0.082 & 0.16 & 2.62 & 67.77 & 0.39 \\
& ESRX & 0.042 & 0.074 & 0.24 & 0.32 & 0.12 & 0.20 & 4.12 & 60.00 & 0.63 \\
& GILD & 0.052 & 0.043 & 0.34 & 0.46 & 0.032 & 0.077 & 1.64 & 48.23 & 0.23 \\  \hline

    \end{tabular}
    \caption{Summary statistics for all stocks, showing the probability of the different events, the mean spread in ticks, the mean price in dollars and the average time 
    between events in seconds. The last column shows the total number of events in the sample.}
    \label{tab:eventprobs}
\end{table*}


\section{Correlation and response functions}
\label{sec:cr}

In this section, we study the empirical temporal correlation of the different events defined above, and the response function to these events.

\subsection{The autocorrelation of $\epsilon$ and $s$}

We first investigate the autocorrelation function of the event signs, calculated as $\ev{\epsilon_{t+\ell}\cdot \epsilon_t}$. These are found to be short-ranged, see Fig. \ref{fig:acsignside}, where the correlation function dies out after 10-100 trades, corresponding to typically 10 seconds in real time. This is in contrast with several other papers \cite{bouchaud.subtle, lillo.farmer, bouchaud.molasses, bouchaud.review}, where $\epsilon$'s are calculated for market orders only ($\pi$ = $\MOO$, $\MO'$), and those signs are known to be strongly persistent \emph{among themselves},
with, as recalled in Sec. \ref{sec:review}, a correlation decaying as a slow power law. However, the direction of incoming limit orders is negatively correlated with cancellations and market orders. Because the $\epsilon$ time series contains all types of events, the mixture of long-range positive and negative correlations balances such that only short-range persistence remains. Any other result would be incompatible with little predictability in price returns. As illustrated by the toy example of Sec. \ref{sec:review}, Eq. (\ref{eq:pttoy3}), this mixing process in fact maintains statistical market efficiency, i.e. weak autocorrelation of price changes.

When limit orders and cancellations are included, one can independently analyze the persistence of the side $s_t$ of the events. According to Eq. \eqref{eq:sign} this means flipping the event signs of limit orders in the $\epsilon$ time series, while keeping the rest unchanged. This change reverses the compensation mechanism discussed above, and $s$ is found to have long-range correlations in time: $\ev{s_{t+\ell} \cdot s_t}$ is shown in Fig. \ref{fig:acsignside} and decays as $\ell^{-\gamma}$ with $\gamma \approx 0.7$. This long range decay is akin to the long range persistence of market order signs discussed throughout the literature: since market orders tend to persistently hit one side of the book, one expects more limit orders and cancellations on the same side as well. Intuitively, if a large player splits his order and buys or sells using market orders for a long period of time, this will attract compensating limit orders on the same side of the book.

\addsmallfig{acsignside}{}{acsignside}{$\ev{\epsilon_{t+\ell}\cdot \epsilon_t}$ and $\ev{s_{t+\ell}\cdot s_t}$, averaged for large and small tick stocks.}

\subsection{The signed event-event correlation functions}

We will see in the following, that for describing price impact the most important correlation functions are those defined between two (not necessarily different) signed event types. For some fixed $\pi_1$ and $\pi_2$ one can define the normalized correlation between these signed events as:
\begin{equation}
    C_{\pi_1, \pi_2}(\ell) = \frac{\ev{I(\pi_{t}=\pi_1)\epsilon_tI(\pi_{t+\ell}=\pi_2)\epsilon_{t+\ell}}}{P(\pi_1)P(\pi_2)}.
    \label{eq:Cpi1pi2}
\end{equation}
Our convention is that the first index corresponds to the first event in chronological order. Because we have $6$ event types, altogether there are $6^2 = 36$ of these event-event correlation functions. There are no clearly apparent, systematic differences between large and small tick stocks, hence we give results averaged over both groups in Fig. \ref{fig:acrenorm_0} for $\pi_1=\MOO$ and
$\pi_1=\MO'$. (Other correlation functions are plotted in Appendix \ref{app:data}.) Trades among themselves and regardless of group are long range correlated as it is well known and was recalled above, and confirmed
again in Fig. \ref{fig:acrenorm_0}. For other cases, the sign of the correlations between event types varies and in many cases one observes a similarly slow decay that can be fitted by a power law with an exponent around $0.5$. Furthermore, there are two distinctly different regimes. For $\ell \lesssim 100$ events (which means up to $10-20$ seconds in real time) returns are still autocorrelated (cf. Fig. \ref{fig:acsignside}). In this regime $C_{\MOO,\pi_2}(\ell)$ is positive for any event type $\pi_2$, so small trades are followed by a ballistic move in the same direction by other trades and also by limit orders, while at the same time cancellations also push the price in the same direction. $C_{\MO',\pi_2}(\ell)$ is also positive except for $\LO'$, where it is negative except for very small lags\footnote{There is some sign of oscillations for small tick stocks.}. This means that if a market
order removes a level, it is followed by further trades and cancellations in the same direction, but the level is refilled very quickly by incoming limit orders inside the spread. For longer times some correlation functions change sign. For example in Fig. \ref{fig:acrenorm_0}(left) one can see this reversal for limit orders. Market orders ``attract'' limit orders, as noted in \cite{weber.rosenow, bouchaud.molasses, gerig.phd}. 
This ``stimulated refill'' process ensures a form of dynamic equilibrium: the correlated flow of market orders is offset by an excess inflow of opposing limit orders, such as to maintain the diffusive nature of the price. This is the same process causing the long-range correlations of $s_t$ noted above.

\addtwofigs{acrenorm_0}{}{acrenorm_all_0}{acrenorm_all_1}{The normalized, signed event correlation functions $C_{\pi_1, \pi_2}(\ell)$, {\it (left)} $\pi_1=\MOO$, {\it (right)} $\pi_1=\MO'$. The curves are labeled by their respective $\pi_2$'s in the legend. The bottom panels show the negative values.}

In general, there are no reasons to expect time reversal symmetry, which would impose $C_{\pi_1, \pi_2}(\ell)=C_{\pi_2, \pi_1}(\ell)$. 
However, some pairs of
events appear to obey this symmetry at least approximately, for example $\MOO$ and $\CAO$ or $\MO'$ and $\CA'$, see Fig. \ref{fig:acrenorm_symmetries_smalltick}. On the other hand, for the pair $\MO'$, $\LO'$  one can see that limit orders that move the price are immediately followed by opposing market orders. The dual compensation, i.e. a stimulated refill of liquidity after a price moving market order $\MO'$, only happens with some delay. $\MOO$ and limit orders also lead to some asymmetry, see Fig. \ref{fig:acrenorm_asymmetries_smalltick}; here we see that after
a transient, non-aggressive market orders induce compensating limit orders more efficiently than the reverse process. 

\addsmallfig{acrenorm_symmetries_smalltick}{}{acrenorm_symmetries_smalltick}{Examples for time reversal symmetry for normalized, signed event correlations for small tick stocks, note that it is $-C_{\pi_1, \pi_2}(\ell)$ plotted. Lines and points of the same color correspond to the same event pairs. The curves are labeled by their respective $\pi_1$'s and $\pi_2$'s in the legend.}

\addtallfig{acrenorm_asymmetries_smalltick}{}{acrenorm_asymmetries2_smalltick}{Examples for time reversal asymmetry for normalized, signed event correlations for small tick stocks. Lines and points of the same color correspond to the same event pairs. The curves are labeled by their respective $\pi_1$'s and $\pi_2$'s in the legend.}

\subsection{The unsigned event-event correlation functions}

A similar definition of a correlation function is possible purely between event occurences, without the signs:
\begin{equation}
\Pi_{\pi_1,\pi_2}(\ell) = \frac{P(\pi_{t+\ell}=\pi_2|\pi_t=\pi_1)}{P(\pi_2)}-1 \equiv \frac{\ev{I(\pi_t=\pi_1)I(\pi_{t+\ell}=\pi_2)}}{P(\pi_1)P(\pi_2)}-1,
\end{equation}
where we have subtracted $1$ such as to make the function decay to zero at large times. This quantity expresses the excess probability of $\pi_2$-type events in comparison to their stationary 
probability, given that there was a $\pi_1$-type event $\ell$ lags earlier. Examples of this quantity for averages over all stocks are plotted in Fig. \ref{fig:acpirenorm_0}. One finds that generally $\Pi_{\pi_1, \pi_2}(\ell)$ decays slower when both $\pi_1$ and $\pi_2$ move the price. 
This implies that events which change the best price are clustered in time: aggressive orders induce and reinforce each other.

\addtwofigs{acpirenorm_0}{}{acpirenorm_all_0}{acpirenorm_all_1}{The normalized, unsigned event correlation functions $\Pi_{\pi_1, \pi_2}(\ell)$, {\it (left)} $\pi_1=\MOO$, {\it (right)} $\pi_1=\MO'$. The curves are labeled by their respective $\pi_2$'s in the legend. The bottom panels show the negative values.}

\subsection{The response function}

Let us now turn to the response of the price to different types of orders. The average behavior of price after events of a particular type $\pi$ defines the corresponding \emph{response function} (or average impact function):
\begin{equation}
    R_\pi(\ell) = \ev{(p_{t+\ell}-p_t) \cdot \epsilon_t |\pi_t = \pi}.
    \label{eq:Rpi}
\end{equation}
This is a correlation function between ``sign times indicator" $\epsilon_t I(\pi_t = \pi)$ at time $t$ and the price change from $t$ to $t+\ell$, normalized by the stationary probability of the event $\pi$, denoted as $P(\pi) = \ev{I(\pi_t = \pi)}$. This normalized response function gives the expected directional price change after an event $\pi$. Its behavior for all $\pi$'s is shown in Fig. \ref{fig:rfnorm}. We note that all type of events lead, on
average, to a price change in the expected direction. Tautologically, $R_\pi(\ell=1) >0$ for price changing events and $R_\pi(\ell=1)=0$ for other
events. As the time lag $\ell$ increases, the impact of market orders grows significantly, specially for small tick stocks, whereas it remains roughly constant for limit orders/cancellations that do change the price. However, as emphasized in \cite{bouchaud.subtle}, the response function is hard
to interpret intuitively, and in particular is not equal to the bare impact of an event since the correlations between events contribute to 
$R_\pi(\ell)$, see Eq. \eqref{eq:RCG} above. We now attempt to deconvolute the effect of correlations and extract these bare impact functions 
from the data.

\addwidefig{rfnorm}{}{R}{The normalized response function $R_\pi(\ell)$ for {\it (left)} large tick stocks and {\it (right)} small tick stocks. The curves are labeled according to $\pi$ in the legend.}

\section{The temporary impact model}
\label{sec:G}

Market orders move prices, but so do cancellations and limit orders. As reviewed in Sec. \ref{sec:review} above, one can try to describe the impact of all these events in an effective way in terms of a ``dressed'' propagator of market orders only, ${G}(\ell)$, as defined by Eq. (\ref{eq:ptMO}). Let us extend this formalism to include any number of events in the following way. We assume, that after a lag of $\ell$ events, an event of type $\pi$ has a 
remaining impact $G_\pi(\ell)$. The price is then expressed as the sum of the impacts of {\it all} past events, plus some initial reference price:
\begin{equation}
    p_t = \sum_{t'<t} G_{\pi_t'}(t-t') \epsilon_{t'} + p_{-\infty},
    \label{eq:pt}
\end{equation}
where the term with the indicators selects exactly one propagator for each $t'$, the one corresponding to the particular event type at that time. After straightforward calculations, the response function \eqref{eq:Rpi} can be expressed through Eq. \eqref{eq:pt} and \eqref{eq:trick2} as
\begin{eqnarray}
    R_{\pi_1}(\ell) = \sum_{\pi_2}P(\pi_2)\left [ \sum_{0 < n\leq \ell} G_{\pi_2}(n) C_{\pi_1, \pi_2}(\ell-n) + 
    \sum_{n>\ell} G_{\pi_2}(n) C_{\pi_2, \pi_1}(n-\ell) - \sum_{n>0} G_{\pi_2}(n) C_{\pi_2, \pi_1}(n)\right].
    \label{eq:undresspi}
\end{eqnarray}
This is a direct extension of Eq. \eqref{eq:RCG}, which was obtained in Ref. \cite{bouchaud.molasses}. One can invert the system of equations in \eqref{eq:undresspi}, to evaluate the unobservable $G_\pi$'s in terms of the observable $R_\pi$'s and $C_{\pi_1, \pi_2}$'s. 
In order to do this, one rewrites the above in a matrix form, as
\begin{eqnarray}
    \label{eq:Gfit}
    R_{\pi_1}(\ell) = \sum_{\pi_2}\sum_{n=0}^\infty A_{\ell,n}^{\pi_1, \pi_2} G_{\pi_2}(n),
\end{eqnarray}
where
\begin{equation}
A_{\ell,n}^{\pi_1, \pi_2} = P(\pi_2) \left\{
\begin{array}{rl}
C_{\pi_1, \pi_2}(\ell-n)-C_{\pi_2, \pi_1}(n), & \text{\ if\ } 0 < n \leq \ell \leq L\\
C_{\pi_2, \pi_1}(n-\ell)-C_{\pi_2, \pi_1}(n), & \text{\ if\ } 0 < \ell < n \leq L\\
\end{array} \right.
\end{equation}
and $\infty$ was replaced by a large enough cutoff $L$, convenient for numerical purposes. In the following, we use $L=1000$, which allows to determine 
the functions $G_\pi$ with a good precision up to $\ell \sim 300$, see Fig. \ref{fig:G}. 

As discussed in Sec. \ref{sec:review}, the origin of the decay of market order price impact is that incoming limit orders maintain an equilibrium with market order flow. In order to keep prices diffusive, limit orders introduce a reverting force into prices, and this precisely off-sets the persistence in market order flow. However, our present extended formalism \emph{explicitly} includes these limit orders (and also cancellations) as events. If 
{\it all} order book events were described, one naively expects that the $G_\pi$'s should be lag-independent constants for events that change the price, and zero otherwise. Solving the above equation for $G_\pi$'s, however, leads to functions that still depend on the lag $\ell$, particularly for small tick stocks: see Fig. \ref{fig:G}. We see in particular that market orders that do not change the price immediately do impact the price on longer time scales. We also
notice that the impact of single $\MO',\MOO$ events first grows with lag and then decays slowly. The impact of limit orders, although clearly measurable, seems to be 
significantly smaller than that of market orders, in particular for small tick stocks (see \cite{hautsch.limit} for a related discussion). 

\addwidefig{G}{}{G}{The bare propagators $G_\pi(\ell)$ in the temporary impact model for {\it (left)} large tick stocks and {\it (right)} small tick stocks.}

In the rest of the paper, we will try to understand in more detail where the lag dependence of $G_\pi$'s comes from. The discussion of Sec. \ref{sec:review} already suggested that some history dependence of impact is responsible for this effect.
Before dwelling into this, it is interesting to see how well the above augmented model predicts the volatility of the stocks once all the $G_\pi$'s have 
been calibrated on the empirical $R_\pi$'s. As just mentioned, Eq. (\ref{eq:pt}) neglects the {\it fluctuations} of the impact, and we therefore expect some discrepancies.  In order to make such a comparison, we first express exactly the variance of the price at lag $\ell$, 
$D(\ell)=\ev{(p_{t+\ell}-p_t)^2}$ in terms of the $G$'s and the $C$'s, generalizing the corresponding result obtained in \cite{bouchaud.subtle}:
\begin{eqnarray}
  \label{eq:Dell}
  D(\ell) &=& \ev{(p_{t+\ell}-p_t)^2} = \sum_{0\leq n < \ell}\sum_{\pi_1} G_{\pi_1}(\ell-n)^2 P(\pi_1) + 
  \sum_{n>0} \sum_{\pi_1} \left [G_{\pi_1}(\ell+n)-G_{\pi_1}(n) \right ]^2 P(\pi_1)  \nonumber \\ 
 &+& 2 \sum_{0\leq n < n' < \ell}\sum_{\pi_1, \pi_2} G_{\pi_1}(\ell-n)G_{\pi_2}(\ell-n')C_{\pi_1,\pi_2}(n'-n)  \nonumber \\
 &+& 2 \sum_{0 < n < n' < \ell}\sum_{\pi_1, \pi_2}\left [G_{\pi_1}(\ell+n)-G_{\pi_1}(n)\right ]\left [G_{\pi_2}(\ell+n')-G_{\pi_2}(n')\right ]C_{\pi_1,\pi_2}(n-n') \nonumber \\
 &+& 2 \sum_{0 \leq n < \ell} \sum_{n' > 0} \sum_{\pi_1, \pi_2} G_{\pi_1}(\ell-n)\left [G_{\pi_2}(\ell+n')-G_{\pi_2}(n')\right ]C_{\pi_2,\pi_1}(n'+n).
\end{eqnarray}
The function $D(\ell)/\ell$, which should be constant for a strictly diffusive process, is plotted in Fig. \ref{fig:vol_test_fair_merged}, the symbols indicate the empirical data, and the dashed lines correspond to Eq. \eqref{eq:Dell}. Note that we fit both models to each stock separately, compute $D(\ell)/\ell$ in each case, and then average the results. We see that the overall agreement is fair for small tick stocks, but very bad for large tick stocks. The reason will turn out to be that for large ticks, a permanent, non fluctuating impact model accounts very well for the dynamics. This reflects that the spread and the gaps behind the best quotes are nearly constant in that case. But any small variation of $G_\pi$ is amplified through the second term of Eq. \eqref{eq:Dell} which is an infinite sum of positive terms. Hence it is much better to work backwards and test a model where the single event propagator is assumed to be strictly constant over time, as we will explain in the next section. 

\addsmallfig{vol_test_fair_merged}{}{vol_test_fair_merged}{$D(\ell)/\ell$ and its approximations for the two groups of stocks. For small tick stocks the values were divided by $10$ for
clarity. Symbols correspond to the empirical result. Dashed lines correspond to the temporary impact model with all $6$ events and they are calculated from Eq. \eqref{eq:Dell}. The agreement is acceptable for small tick stocks, but very poor for large tick ones. Solid lines correspond to the constant impact model, see Eq. (\ref{eq:permanent2}) below; in this case the agreement with large tick stocks in nearly perfect, at least visually.}

\section{A constant impact model}
\label{sec:perm}

In the above section we found that the single event propagators $G_\pi$ appear to have a non-trivial time dependence. Another way to test this 
result is to invert the logic and assume first that the $G_\pi$ are time {\it independent} and see how well, or how badly, this assumption fares at accounting
for the shape of the response functions $R_\pi(\ell)$ and of the price diffusion $D(\ell)$.

Let us start from the following \emph{exact} formula for the midpoint price: 
\begin{equation}
    \label{eq:permanent}
    p_{t+\ell} = p_t + \sum_{t\leq t'< t+\ell} \epsilon_{t'}  \Delta_{\pi_{t'}, \epsilon_{t'}, t'} .
\end{equation}
Here $\Delta_{\pi, \epsilon_{t'}, t'}$ denotes the price change at time $t'$ if an event of type $\pi$ happens. This $\Delta$ can also depend on the sign $\epsilon_{t'}$. For example, 
if $\pi=\MO'$ and $\epsilon_{t'}=-1$ this means that at a sell market order executed the total volume at the bid. The midquote price change is $-\Delta_{\MO', -1, t'}$, which usually means that the
second best level was $b_{t'}-2\Delta_{\MO', -1, t'}$, where $b_{t'}$ is the bid price before the event. The factor $2$ is necessary, because the ask
did not change, and the impact is defined by the change of the midquote. Hence $\Delta_{\MO'}$'s (and similarly $\Delta_{\CA'}$'s) correspond to half of the gap between the first and the second best quote \emph{just before} the level was removed (see also Ref. \cite{farmer.whatreally}). Another example when $\pi=\LO'$ and $\epsilon_{t'}=-1$. This means that at $t'$ a sell limit order was placed inside the spread. The midquote price change is $-\Delta_{\LO', -1, t'}$, which means that the limit order was placed at $a_{t'}-2\Delta_{\LO', -1, t'}$, where $a_{t'}$ is the ask price. Thus $\Delta_{\LO'}$'s correspond to half of the gap between the first and the second best quote \emph{right after} the limit order was placed. In the following we will call the $\Delta$'s \emph{gaps}. Note that the events $\MOO$, $\CAO$ and $\LOO$ do not change the price, so their respective gaps are always zero: there are only three types of $\Delta$'s that are non-zero.

The permanent impact model is defined by replacing the time dependent $\Delta$'s by their average values. More precisely, let us introduce the
average realized gap:
\begin{equation}
\label{eq:deltaR}
  \DeltaR_{\pi} = \ev{\Delta_{\pi_t, \epsilon_{t}, t}|\pi_{t}=\pi}.
\end{equation}
The conditional expectation means that the gaps are sampled only when the price change corresponding
to that particular kind of gap is truly realized. Therefore, in general $\Delta^\mathrm{R}_{\pi} \neq \ev{\Delta_{\pi, \epsilon_{t}, t}}$, see Table \ref{tab:gaps} where
one sees that the realized gap when a market order moves the price is in fact \emph{larger} than the unconditional average. The logic is that the opening of a 
large gap behind the ask is a motivation for buying rapidly (or cancelling rapidly for sellers) before the price moves up.

Our approximate {\it constant impact} model then reads:
\begin{equation}
   \label{eq:permanent2}
    p_{t+\ell} = p_t + \sum_{t\leq t'< t+\ell} \Delta^\mathrm{R}_{\pi_{t'}} \epsilon_{t'}.
\end{equation}
The response functions are then, by using Eq. \eqref{eq:trick2}, easily given by:
\begin{eqnarray}
R_{\pi}(\ell) = \ev{(p_{t+\ell}-p_t) \cdot \epsilon_t |\pi_t = \pi} = \sum_{0\leq t'< \ell} \sum_{\pi_1} \Delta^\mathrm{R}_{\pi_1} P(\pi_1) C_{\pi, \pi_1}(t'),
    \label{eq:rmean}
\end{eqnarray}
The formula \eqref{eq:rmean} is quite simple to interpret. We fixed that the event that happened at $t$ was of type $\pi$. Let us now 
express $C_{\pi, \pi_1}(\ell)$ as:
\begin{eqnarray} 
P(\pi)P(\pi_1)C_{\pi, \pi_1}(\ell) \propto P(\pi_{t+\ell}=\pi_1, \epsilon_{t+\ell} = \epsilon_{t}|\pi_{t}=\pi) - P(\pi_{t+\ell}=\pi_1, \epsilon_{t+\ell}=-\epsilon_{t}|\pi_{t}=\pi).
\end{eqnarray}
This represents the following: Given that the event at $t$ was of type $\pi$ and the event at $t+\ell$ is of type $\pi_1$, how much more is it probable, that the direction of the second event is the same as that of the first event? The total price response to some event can be understood as its own impact (lag zero), plus the sum of the biases in the course of future events, conditional to this initial event. These biases are multiplied by the average price change $\Delta^\mathrm{R}$ that these induced future events cause. Of course, correlation does not mean causality, and we cannot a priori distinguish between events that are {\it induced} by the initial event, and those that merely {\it follow} the initial event (see \cite{farmer.zamani} 
for a related discussion). However, it seems 
reasonable to assume that there is a true causality chain between different types of events occuring on the same side of the book (i.e. a limit order
refilling the best quote after a market order).

\begin{table*}[tbp]
    \begin{tabular}{|c|c||c|c|c|c|}
	\hline
& ticker & $2\DeltaR_{\MO'}$ & $2\DeltaR_{\CA'}$ & $2\DeltaR_{\LO'}$ & $2\ev{\Delta_{\MO'}}$ \\ \hline\hline
\multirow{7}{*}{\begin{sideways}large tick\end{sideways}} & AMAT & 1.02 & 1.04 & 1.02 & 1.00 \\
& CMCSA & 1.03 & 1.14 & 1.06 & 1.00 \\
& CSCO & 1.01 & 1.02 & 1.01 & 1.00 \\
& DELL & 1.01 & 1.05 & 1.02 & 1.00 \\
& INTC & 1.00 & 1.01 & 1.01 & 1.00 \\
& MSFT & 1.01 & 1.02 & 1.01 & 1.00 \\
& ORCL & 1.01 & 1.02 & 1.02 & 1.00 \\ \hline

\multirow{6}{*}{\begin{sideways}small tick\end{sideways}} & AAPL & 1.31 & 1.27 & 1.27 & 1.14 \\
& AMZN & 1.51 & 1.22 & 1.30 & 1.17 \\
& APOL & 1.76 & 1.50 & 1.52 & 1.42 \\
& COST & 1.35 & 1.23 & 1.24 & 1.15 \\
& ESRX & 1.85 & 1.54 & 1.60 & 1.45 \\
& GILD & 1.11 & 1.13 & 1.11 & 1.03 \\ \hline

    \end{tabular}
    \caption{Mean realized gaps and unconditional gaps in ticks for all stocks. All values were multiplied by $2$, so that they correspond to the instantaneous change of the bid/ask and not of the midquote. Note that $\ev{\Delta_{\MO'}}=\ev{\Delta_{\CA'}}$, while $\ev{\Delta_{\LO'}}$ is
    not observable.}
    \label{tab:gaps}
\end{table*}

Let us now take Eq. \eqref{eq:rmean}, and check how well the true response functions are described by the above constant impact model. Figs. \ref{fig:redressepstest023} and \ref{fig:redressepstest145} show that the agreement is very satisfactory for large tick stocks, except when $\pi_1=\CA'$, but these events are very rare (less than $\sim 0.2\%$). This agreement is expected because the order book is usually so dense that gaps hardly fluctuate at all; the small remaining discrepancies will in fact be cured below. The quality of the agreement suggests that the time dependence of 
the bare impact function $G_\pi$ obtained in Sec. \ref{sec:G} above is partly a numerical artefact coming from the ``brute force'' inversion of Eq. (\ref{eq:Gfit}). 

For small ticks on the other hand, noticeable deviations are observed as expected, and call for an extension of the model. This will be the focus of the next sections. One can extend the above model in yet another direction, by studying the dynamics of the spread rather than the dynamics of the mid-point, see Appendix \ref{app:spread}.

\addwidefig{redressepstest023}{}{redressepstest023}{Comparison of true and approximated normalized response functions $R_\pi(\ell)$,
using the constant gap model, for {\it (left)} large tick stocks and {\it (right)} small tick stocks, for events that {\it do not} change the price. Symbols correspond to the true value, and lines to the approximation. The data are labeled according to $\pi$ in the legend.}
\addwidefig{redressepstest145}{}{redressepstest145}{Comparison of true and approximated normalized response functions $R_\pi(\ell)$,
using the constant gap model, for {\it (left)} large tick stocks and {\it (right)} small tick stocks, for events that change the price. Symbols correspond to the true value, and lines to the approximation. The data are labeled according to $\pi$ in the legend.}

One can approximate the volatility within the same model as
\begin{eqnarray}
    D(\ell) = \ev{(p_{t+\ell}-p_t)^2} \approx \sum_{0\leq t', t''<\ell} \sum_{\pi_1} \sum_{\pi_2} P(\pi_1) P(\pi_2) 
    C_{\pi_1, \pi_2}(t'-t'') \Delta^\mathrm{R}_{\pi_1} \Delta^\mathrm{R}_{\pi_2}.
    \label{eq:vol}
\end{eqnarray}
As shown in Fig. \ref{fig:vol_test_fair_merged}, the constant gap model is very precise for large tick stocks (as again expected), but clear discrepancies are
visible for small tick ones.

\section{The gap dynamics of small tick stocks}
\label{sec:gap}
\subsection{A linear model for gap fluctuations}

Let us now try to better understand how gap fluctuations contribute to the response function, and why replacing the gap by its average realized value is not a good approximation for small tick stocks. By definition, without the constant gap approximation, the response function contains contributions which have the form
$$\ev{\Delta_{\pi_2, \epsilon_{t+\ell}, t+\ell}\epsilon_{t+\ell}\epsilon_t\bigg|\pi_{t}=\pi_1,\pi_{t+\ell}=\pi_2}.$$ After using some basic properties of the event signs this quantity can be written as a sum over three contributions:
\begin{enumerate}
  \item Firstly, there is the term from the constant gap approximation:
  $$\DeltaR_{\pi_2}\ev{\epsilon_t\epsilon_{t+\ell}\bigg|\pi_{t}=\pi_1,\pi_{t+\ell}=\pi_2}.$$
  This contains the highest order of the effect of event-event correlations.
  \item There is a second term that we write as:
  $$\frac{1}{2}\ev{\underbrace{[\Delta_{\pi_2, +, t+\ell}-\Delta_{\pi_2, -, t+\ell}]\epsilon_t}_{(a)}\bigg|\pi_{t}=\pi_1,\pi_{t+\ell}=\pi_2},$$
  which is the conditional expectation value of the quantity $(a)$. If $(a)$ is positive, then after an upward price move consecutive upward moves are larger than downward ones, while if $(a)$ is negative then they are smaller. This process can thus either accelerate or dampen the growth of the response function.
  \item The third contribution is of the form
  $$\frac{1}{2}\ev{\underbrace{[\Delta_{\pi_2, +, t+\ell}+\Delta_{\pi_2, -, t+\ell}-2\DeltaR_{\pi_2}]}_{(b)}\underbrace{\epsilon_{t}\epsilon_{t+\ell}}_{(c)} \bigg|\pi_{t}=\pi_1,\pi_{t+\ell}=\pi_2}.$$
  Here $(b)$ is positive, when the average of the two gaps (up and down) is greater than the time averaged realized value. $(c)$ is positive, when the two events move the price in the same direction. Thus the full term gives a positive contribution to the response function, if two ``parallel" events are correlated with larger gaps and hence decreased liquidity at the time of the second event, while opposing events correspond to increased liquidity at the time of the second event. The final effect of this term agrees with the previous one: If $(b)\times (c)$ is positive, then after an upward price move the consecutive upward moves become larger than downward ones and vice versa.
\end{enumerate}

At this point we need a dynamical model for the $\Delta$'s, to quantify the above correlations, but we are faced with the difficulty that $\Delta_{\pi, \epsilon, t}$ is only observed for $\pi=\pi_t$ and $\epsilon=\epsilon_t$. What we will do instead is to write a simple regression model directly for the observable quantity $\Delta_{\pi, \epsilon_t, t}I(\pi_t=\pi)\epsilon_t$, that can be evaluated from data. Then based on this knowledge we will revisit the influence of gap fluctuations on the price dynamics in Sec. \ref{sec:gap2}.


\subsection{A linear model for gap fluctuations}
\label{sec:gapbare}
The correlation between events has a dynamical origin: market orders and cancellations attract replacement limit orders and vice versa. Eq. \eqref{eq:permanent} is the exact time evolution of price written as a sum of the random variables $\Delta_{\pi, \epsilon_{t}, t} I(\pi_{t}=\pi)\epsilon_{t}$. We will postulate that both the realized gap $\Delta_{\pi, \epsilon_t, t}$ and the order flow 
$I(\pi_t=\pi)\epsilon_t$ are influenced by the past order flow $I(\pi_{t'}=\pi)\epsilon_{t'}$, $t' < t$ in a linear fashion, i.e.:
\begin{equation}
\Delta_{\pi, \epsilon_{t}, t} I(\pi_{t}=\pi)\epsilon_{t} = \sum_{t'<t} \sum_{\pi_1} K_{\pi_1, \pi}(t-t')I(\pi_{t'}=\pi_1)\epsilon_{t'}+\eta_{\pi_1, t},
\label{eq:Knew}
\end{equation}
where all $\eta$'s are independent noise variables. Similarly, we write for the three price changing events $\MO',\LO'$ and $\CA'$:
\begin{equation}
\DeltaR_{\pi} I(\pi_{t}=\pi)\epsilon_{t} = \sum_{t'<t} \sum_{\pi_1} \tilde K_{\pi_1, \pi}(t-t')I(\pi_{t'}=\pi_1)\epsilon_{t'}+\tilde\eta_{\pi_1, t},
\label{eq:Knew2}
\end{equation}
with other noise variables $\tilde \eta$, and we introduced $\DeltaR_{\pi}$ for later convenience. Note the above equations are again of the vector autoregression type, where the kernel 
$K$ and $\tilde K$ have a $3 \times 6$ matrix structure.  

Both models \eqref{eq:Knew} and \eqref{eq:Knew2} can be calibrated to the data by using the same trick an in Sec. \ref{sec:G}, forming expectation values on both sides and solving a set of linear equations between correlation functions, for example for $K$:
\begin{equation}
\ev{\Delta_{\pi, \epsilon_{t+\ell}, t+\ell} I(\pi_{t+\ell}=\pi)\epsilon_{t+\ell}I(\pi_{t}=\pi_1)\epsilon_{t}} = \sum_{t'<t+\ell} \sum_{\pi_2} K_{\pi_2, \pi}(t+\ell-t')\ev{I(\pi_{t'}=\pi_2)\epsilon_{t'}I(\pi_{t}=\pi_1)\epsilon_{t}},
\end{equation}
except this time we have three separate solutions for $\pi=\MO'$, $\CA'$ and $\LO'$. An example of the solution kernels $K$ is given in Fig. \ref{fig:undressD_after_realized_1_smalltick}; the sign of these kernels is expected from what we learnt in Sec. \ref{sec:cr}. We see for example that
a $\MO$ event tends to make a future $\MO'$ more probable, and with an increased gap, which makes sense. The same can be repeated with respect to Eq. \eqref{eq:Knew2} to calculate the $\tilde K$'s.

\addsmallfig{undressD_after_realized_1_smalltick}{}{undressD_after_realized_1_smalltick}{Estimates of $K_{\pi, \MO'}(t)$ for small ticks.}

An important aspect of these VAR models is that once we have an estimate for their kernels, they can be used for forecasting the future price changes caused by each component of the event flow based on the events that occured in the recent past \cite{hasbrouck.book}. Eq. \eqref{eq:Knew} prescribes for us an estimate for conditional expectation values such as $\ev{\Delta_{\MO', \epsilon_{t}, t} I(\pi_{t}=\MO')\epsilon_{t}|\cdots}$, which is the expected price change due to a market order in the next event (times the probability of such an outcome), and the conditioning is on past signs and indicators. We can proceed similarly for $\CA'$ and $\LO'$, and finally the sum of the three components gives the expected price change in the next event.

Such forecasts based on Eqs. \eqref{eq:Knew}, \eqref{eq:Knew2} perform surprisingly well in practice, although liquidity is fragmented and some events are unobserved. Fig. \ref{fig:undressD_after_realized_run_smalltick} shows that the expectation value of the left hand side of Eq. \eqref{eq:Knew} is a monotonic function of our prediction, and the relationship on average can be fitted with a straight line with slope $1$, although small higher order (cubic) corrections seem to be present as well. Similar results can be found for Eq. \eqref{eq:Knew2} and $\tilde K$'s.

As discussed in Sec. \ref{sec:review}, one should interpret the kernels $K$ as ``dressed'' objects that include the contribution of events that occur on unobserved platforms. This is justified as
long as one is concerned with linear observables, such as average response functions. For non-linear quantities, such as diffusion, some discrepancies are expected.

\addsmallfig{undressD_after_realized_run_smalltick}{}{undressD_after_realized_run_smalltick}{Performance of Eq. \eqref{eq:Knew} for small ticks. Both axes normalized by standard deviation of predictor.}

\subsection{The final model for small ticks}
\label{sec:gap2}

The above analysis suggests a way to build and calibrate an impact model that describes in a consistent way (a) all types of events and (b) the history dependence of the gaps, as we argued to be necessary in Sec. \ref{sec:review}. The discussion of the previous section motivates the following model: 
\begin{equation}
\label{eq:finalmodel}
  p_{t+\ell} = p_t + \sum_{t\leq t'< t+\ell} \left [\Delta^R_{\pi_{t'}}+  \sum_{t''<t'}\kappa_{\pi_{t''},\pi_{t'}}(t'-t'')\epsilon_{t'} \epsilon_{t''}\right] \epsilon_{t'},
\end{equation}
where $\kappa_{\pi_2,\pi_1}$ is a kernel that models the fluctuations of the gaps and their history dependence, which will be chosen such that the
bare propagator of the model is given by Eq. (\ref{eq:Gfinal}) above. 

The model specification, Eq. \eqref{eq:finalmodel}, is the central result of this paper. It can be seen as a {\it permanent impact model}, but with some history dependence, modeled as a linear regression on past events. By symmetry, this dependence should only include terms containing $\epsilon_{t'} \epsilon_{t''}$ since the influence of any past string of events on the ask must be the same as that of the mirror image of the string on the bid. More generally, one may expect higher order, non-linear correction terms of the form
\begin{equation}
\sum_{t_1,t_2,t_3 <t'}\kappa_{\pi_{t_1},\pi_{t_2},\pi_{t_3};\pi_{t'}}(t'-t_1,t'-t_2,t'-t_3)\epsilon_{t_1} \epsilon_{t_2} \epsilon_{t_3}\epsilon_{t'},
\end{equation}
or with a larger (even) number of $\epsilon$'s. We will not explore such corrections further here, although Fig. \ref{fig:undressD_after_realized_run_smalltick} suggests these terms are present.

Upon direct identification of Eq. \eqref{eq:finalmodel} with Eq. \eqref{eq:permanent}, and using \eqref{eq:Knew} and \eqref{eq:Knew2} one finds that $\kappa$ can be expressed in terms of 
$K$ and $\tilde K$ as:
\begin{equation}
\label{eq:finalid}
    \kappa_{\pi,\pi_2}(\ell) = K_{\pi,\pi_2}(\ell)-\tilde K_{\pi,\pi_2}(\ell).
\end{equation}
We can now compute the average response functions $R_\pi(\ell)$ and the diffusion curve $D(\ell)$ within this model, and compare the results
with empirical data.

\addwidefig{redress_redressD_test023}{}{errorbars_023}{Comparison of true and approximated normalized response functions $R_\pi(\ell)$ of the final model for {\it (left)} large tick stocks and {\it (right)} small tick stocks, for events that do not change the price. Symbols correspond to the true value, and lines to the approximation. To illustrate the goodness of fit on a stock by stock basis, we calculated the absolute difference between the true and the approximated value, the average of this quantity across stocks is indicated by the error bars. The inaccuracy for large $\ell$ is due to a finite size effect in matrix inversion. The data are labeled according to $\pi$ in the legend.}
\addwidefig{redress_redressD_test145}{}{errorbars_145}{Comparison of true and approximated normalized response functions $R_\pi(\ell)$ of the final model for {\it (left)} large tick stocks and {\it (right)} small tick stocks, for events that change the price. The inaccuracy for large $\ell$ is due to a finite size effect in matrix inversion. Symbols correspond to the true value, and lines to the approximation. To illustrate the goodness of fit on a stock by stock basis, we calculated the absolute difference between the true and the approximated value, the average of this quantity across stocks is indicated by the error bars. The data are labeled according to $\pi$ in the legend.}

For the response functions, the addition of the fluctuating gap term in Eq. \eqref{eq:finalmodel} corrects the small discrepancies found within the constant impact model for large tick stocks. It also allows one to capture very satisfactorily the response function for small tick stocks, see Figs. \ref{fig:redress_redressD_test023} and \ref{fig:redress_redressD_test145}.\footnote{Note that in making these plots we neglected the first $30$ and last $40$ minutes of trading days, so they slightly differ from those in Sec. \ref{sec:perm}. The results of the constant gap model are essentially unchanged regardless of such an exclusion.}

A much more stringent test of the model is to check the behaviour of the diffusion curve $D(\ell)$. The exact calculation in fact involves three and four-point correlation functions, for which we have 
no model. A closure scheme where these higher correlation functions are assumed to factorize yields the following approximation: 
\begin{eqnarray}
  D(\ell) = \ev{(p_{t+\ell}-p_t)^2} \approx \sum_{0\leq t', t''<\ell} \sum_{\pi_1} \sum_{\pi_2} P(\pi_1) P(\pi_2) 
  C_{\pi_1, \pi_2}(t'-t'') \Delta^\mathrm{R}_{\pi_1} \Delta^\mathrm{R}_{\pi_2} + \nonumber \\
  2 \sum_{-\ell < t < \ell} \sum_{\pi_2, \pi_3} \sum_{\tau > 0} (\ell - |t|) \DeltaR_{\pi_3} \kappa^+_{\pi_2, \pi_3} (\tau, t) C_{\pi_2, \pi_3}(t+\tau) P(\pi_2)P(\pi_3) + \nonumber \\
  \sum_{-\ell < t < \ell} \sum_{\pi_2, \pi_4}\sum_{\tau, \tau'>0} (\ell - |t|) \kappa^{++}_{\pi_2,\pi_4}(\tau, \tau',t)C_{\pi_2, \pi_4}(\tau-\tau'+t)P(\pi_2)P(\pi_4),
  \label{eq:Dellhybrid2}
\end{eqnarray}
where
\begin{eqnarray}
  \kappa^+_{\pi_2, \pi_3}(\tau, t) = \sum_{\pi_1} \kappa_{\pi_2, \pi_1}(\tau)[I(t=0)I(\pi_1=\pi_3)+I(t\not = 0)P(\pi_1)+
  I(t = -\tau) P(\pi_1) \Pi_{\pi_2 \pi_1} (\tau)],
\end{eqnarray}
and, for $t \geq 0$, 
\begin{eqnarray}
  \kappa^{++}_{\pi_2,\pi_4}(\tau, \tau', t) = \sum_{\pi_1, \pi_3} \kappa_{\pi_2, \pi_1}(\tau)\kappa_{\pi_4, \pi_3}(\tau')\{I(t=\tau')I(\pi_1=\pi_4)P(\pi_3) + \nonumber \\ I(t\not=\tau')P(\pi_1)P(\pi_3)[\Pi_{\pi_1, \pi_3}(t)+1]\},
\end{eqnarray}
whereas for $t<0$, we use $\kappa^{++}_{\pi_2,\pi_4}(\tau, \tau', -t) = \kappa^{++}_{\pi_4,\pi_2}(\tau', \tau, t)$. Direct numerical simulation of Eq. \eqref{eq:finalmodel} confirms that our approximation yields $D(\ell)$ curves which are indistinguishable from those of the true model. As Fig. \ref{fig:vol_test_fair_hybrid2const} shows, for small tick stocks this approximation indeed shows some improvement for large $\ell$ when compared to the constant gap model. For small $\ell$ there is still a discrepancy coming from errors in the data that adds some spurious high frequency white noise. To account for these, we add an effective, lag-independent constant to $D(\ell)$, whose value was chosen as $D_0=0.04\mathrm{\ ticks\ squared}$. According to Fig. \ref{fig:vol_test_fair_hybrid2const}, this substantially improves the fit for short times, while leaving the long time contribution unaffected. 

The conclusion is that our history dependent impact model reproduces the empirical average response function in a rather accurate way, and also improves the estimation of the diffusion curve. The discrepancies are expected, since we have neglected several effects, including (i) all volume dependence, (ii) unobserved events deeper in the book and on other platforms and (iii) higher order, non-linear contributions to model history dependence.

\addsmallfig{vol_test_fair_hybrid2const}{}{vol_test_fair_hybrid2const}{$D(\ell)/\ell$ and its approximations for small tick stocks. Symbols correspond to the true result excluding the beginning and the end of trading days, the red line corresponds to the permanent impact model with constant gaps, the green line to the fluctuating gap model (both analytically and by simulation), and the blue line to the fluctuating gap model plus constant. {\it Note:} Unlike the small tick data in Fig. \ref{fig:vol_test_fair_merged}, the vertical axis was not rescaled here.}

\subsection{Interpretation: direct impact vs. induced impact}

The relevance of the history kernels $K$ and $\tilde K$ can be understood through a simple argument. The sequence of events is characterized by the time series $\{\pi_t, \epsilon_t\}$ and together with the gaps this series defines the course of the price. How will the event $\pi$ at time $t$ affect the price at some later time $t+\ell$? This quantity, denoted by $G^*_\pi(\ell)$, is  defined as the average of the formal {\it total derivative} of the price with respect to the past order flow:
\begin{equation}
G^*_\pi (\ell) = \left\langle \frac{d p_{t+\ell}}{d\xi_t^\pi} \right\rangle, \qquad \xi_t^\pi \equiv I(\pi_t=\pi) \epsilon_t.
\end{equation}
It contains two distinct contributions:
\begin{itemize}
  \item A direct one: the immediate price change caused by the event, which is {\it constant in time} (zero or non-zero depending on the value of the corresponding gap). For example, if right now a large buy market order is submitted, it will cause an immediate upward jump in the (ask) price. The average of such jumps due to events of type $\pi$ is represented by the mean realized gap $\DeltaR_\pi$.
  \item An induced, dynamic one: the change of the future event rates and their associated gaps. This modeled by Eq. \eqref{eq:Knew}, and quantified by the kernels $K$. To continue the above example of a large market order, it removes the best ask level, and hence we move into a denser part of the order book. The new first gap behind the ask is on average smaller here. So in effect, our initial event makes the ask gap shrink. In addition, some time after we submit our order, additional sell limit orders will arrive to compensate part of our upwards price pressure. These may move the (ask) price back downwards. If we had decided not to submit our market order, these extra limit orders would not arrive either.
\end{itemize}

To put this decomposition into precise quantitative terms, recall that as an exact identity, 
\begin{equation}
  p_{t+\ell}=p_t +  \sum_{t\leq t'<t+\ell}\sum_{\pi_1} \Delta_{\pi_1, \epsilon_{t'}, t'} \xi_{t'}^{\pi_1}.
\end{equation}
The average derivative of the price with respect to an earlier event is therefore:
\begin{eqnarray}
  G^*_\pi (\ell) =  \sum_{t\leq t'<t+\ell}\sum_{\pi_1}\ev{\frac{d[\Delta_{\pi_1, \epsilon_{t'}, t'}\xi_{t'}^{\pi_1}]}{d\xi_{t}^{\pi}}} = \nonumber \\
  \underbrace{\sum_{\pi_1}\ev{\frac{d[\Delta_{\pi_1, \epsilon_{t}, t}\xi_{t}^{\pi_1}]}{d\xi_{t}^{\pi}}}}_{\DeltaR_\pi} + \sum_{t< t'<t+\ell}\sum_{\pi_1}\underbrace{\left[\ev{\frac{d[\Delta_{\pi_1, \epsilon_{t'}, t'}]}{d\xi_{t}^{\pi}}} \xi_{t'}^{\pi_1} +\underbrace{\DeltaR_{\pi_1}\ev{\xi_{t'}^{\pi_1}}{d\xi_{t}^{\pi}}}_{\tilde K_{\pi, \pi_1}(t'-t)}\right]}_{K_{\pi, \pi_1}(t'-t)}
\end{eqnarray}
where we have introduced the formal definitions of the kernels $K$ and $\tilde K$, in a way compatible with the linear model specified by Eqs. \eqref{eq:Knew} and \eqref{eq:Knew2}.
Therefore, the average price change until time $t+\ell$ attributed to an event of type $\pi$ at time $t$ is found to be:
\begin{eqnarray}
  G^*_\pi (\ell) = \Delta^R_\pi + \sum_{0<t'<\ell}\sum_{\pi_1} K_{\pi,\pi_1}(t').
 \label{eq:Gstar2}
\end{eqnarray}
Numerically, the $G_\pi^*$'s are given in Fig. \ref{fig:redress_redressD_both_after_realized_smalltick}(left) for small tick stocks, 
very similar curves were found for large ticks but we will not detail those here. 

These $G^*$'s are, however, different from the bare response functions $G_\pi$ which are defined as a {\it partial derivative} of the price with respect to event flow (see Eqs. \eqref{eq:Gdef}, 
\eqref{eq:pt}), where all events except the one occuring at $t$ are kept fixed: 
\begin{equation}
  G_\pi(\ell) = \left\langle \frac{\partial p_{t+\ell}}{\partial \xi_t^\pi} \right\rangle.
  \label{eq:Gstarlike}
\end{equation}
Following the logic of the previous calculation and reindexing the terms, one finds, within the linear model 
\begin{eqnarray}
  G_\pi (\ell) = \DeltaR_\pi + \sum_{t<t'<t+\ell}\sum_{\pi_1}\ev{\frac{\partial \Delta_{\pi_1, \epsilon_{t'}, t'}}{\partial \xi_t^\pi}\xi_{t'}^{\pi_1}} = \DeltaR_\pi + \sum_{0<t'<\ell}\sum_{\pi_1} \left[K_{\pi, \pi_1}(t')-\tilde K_{\pi, \pi_1}(t')\right].
  \label{eq:Gstarlike2}
\end{eqnarray}
What is the difference between Eq. \eqref{eq:Gstar2} and Eq. \eqref{eq:Gstarlike2}? In the former, we calculate the total price change until time $t+\ell$ due to the initial event, and this includes the adaptation of future event flow and of future gaps. In the latter, we only keep the possible jump due to this event and the adaptation of gaps, but \emph{not} of the event flow, which is assumed to be fixed. This omission is indeed consistent with Eq. \eqref{eq:pt}, since the effect of event flow adaptation is already accounted for: the equation is based on the true event flow, and hence already includes the full correlation structure between events.

When the tick is small, the gaps are allowed to fluctuate and adapt to the order flow. The extra impact contribution is therefore captured by the above term:
\begin{equation}
  \delta G^*_\pi (\ell) = \sum_{0<t<\ell}\sum_{\pi_1} \left[K_{\pi,\pi_1}(t)-\tilde K_{\pi,\pi_1}(t)\right].
  \label{eq:deltaGstar}
\end{equation}
Our final model defined by Eqs. \eqref{eq:finalmodel} and \eqref{eq:finalid} amounts to adding this fluctuating gap contribution to the average realized gap in the bare propagator, i.e.,
\begin{equation}
G_{\pi}(\ell) = \Delta^R_\pi + \delta G^*_\pi (\ell).
\label{eq:Gfinal}
\end{equation}
The new second term describes the contribution of the gap ``compressibility'' to the impact of an event up to a time lag $\ell$, and it is shown in Fig. \ref{fig:redress_redressD_both_after_realized_smalltick}(right). Perhaps surprisingly, it appears that a small market order $\MOO$ ``softens" the book for small ticks: the gaps tend to grow on average and $\delta G^*_{\MOO}$ is positive. Price changing events on the other hand ``harden" the book, for all stocks the contribution is negative. Queue fluctuations ($\CAO$ and $\LOO$) seem less important, but for small ticks these types of events also harden the book. For large ticks $\delta G^*$'s are found to be about two orders of magnitude smaller, which confirms that gap fluctuations can be neglected to a good approximation in that case.

\addwidefig{redress_redressD_both_after_realized_smalltick}{}{redress_redressD_both_after_realized_smalltick}{{\it (left)} Final estimate of the total average price change $G^*_\pi(\ell)$ due to an event $\pi$, based on Eq. \eqref{eq:Gstar2}, for small tick stocks. {\it (right)} Contribution of gap flexibility to the price change: $\delta G^*_\pi(\ell)$ calculated from Eq. \eqref{eq:deltaGstar} for small tick stocks. The curves are labeled according to $\pi$ in the legend.}

\section{Conclusions}

Previous studies have focused on the impact of market orders only and have concluded that this impact decays in such a way to offset the correlation of the 
sign of the trades. The underlying mechanism is that market orders on one side of the book attract compensating limit orders. These limit orders do not necessarily change the 
best limits, but are such that the conditional impact of a buy trade following other buy trades is smaller than the conditional impact of a sell trade following buy trades.
As pointed out in Gerig \cite{gerig.phd}, the strength of this asymmetric liquidity effect is the dominant effect that mitigates persistent trends in prices. Our study confirms this finding: events happening on the same side of the book are long-range correlated, but the signed correlation function (that assigns an opposite sign to limit order and market order on the same side of the book) is short ranged, 
demonstrating the compensating effect alluded to above.

This effect leads to a strong ``dressing'' of the bare impact of market orders by limit orders. In fact, by including -- besides the market orders -- all limit orders and cancellations at the bid/ask, the price becomes a pure jump process. Every price change, whatever its cause (news, information or noise), can be attributed to exactly one of these events. In a first approximation, the various event types 
lead to a constant jump size that equals the average price change they cause. This simple picture works very well for large tick stocks, where both the average impact of all event types, and the volatility, are quantitatively reproduced by a constant jump model. The situation is different for small tick stocks, where the history dependence of these otherwise permanent jumps becomes important. Note that the effect discussed here is related to, but different from the Lillo-Farmer model that connects the temporal decay of the dressed market order impact to the history dependent conditional impact of a new trade. Here, we are speaking of the history dependence in a framework where the impact of all events, not only of market orders, is already accounted for. 

Another important observation is that not only the jump sizes are history dependent, the events themselves also behave in an adaptive way. An event can induce further events that amplify or dampen its own effect. As it is well known, the arrival of excess buy market orders is shortly followed by additional sell limit orders, but this is just one manifestation of such adaptive dynamics. For example, the reverse process, i.e. market orders following an excess of limit orders are also present, albeit with some delay and a smaller intensity. Our description of these and similar mechanisms, also involving cancellations, is a generalization of the theory of market order price impact in the related literature.

In sum, the dynamics of prices consists of three processes: instantaneous jumps due to events, events inducing further events and thereby affecting the future jump {\it probabilities} (described by the correlation between events), and events exerting pressure on the gaps behind the best price and thereby affecting the future jump {\it sizes}. By approximating this third effect with a linear regression process, we have written down an explicit model, Eq. \eqref{eq:finalmodel}, that accounts very satisfactorily for most of our observations. We have shown how to calibrate such a model on empirical data using some auxiliary kernels $K$ and $\tilde K$ defined by Eqs. \eqref{eq:Knew} and \eqref{eq:Knew2}. This way of extracting the bare propagator $G_\pi$,
motivated by the above decomposition, seems to be less prone to numerical errors than the ``brute force'' inversion method used in Sec. \ref{sec:G}.

The methods proposed in this work are rather simple and general, and can be adapted to measure the impact of any type of trade once a discrete categorization is adopted. One could for example subdivide the category $\MOO$ into small volumes and large volumes, or look at the impact of different option trades on the underlying, etc. Here, we have established that the bare impact of market orders is clearly larger than that of limit orders and that the bare impact of price-changing events shows only partial decay on the time scale that we are able to probe ($1000$ events only corresponds to a few minutes). It would certainly be interesting to study the long time behavior of these bare impact functions, as well as to understand how these impact functions behave overnight.

We hope to have provided here a consistent and complete framework to describe price fluctuations and impact at the finest possible scale. Our approach can be seen as a ``microscopic'' 
construction of VAR-like models, with a clearly motivated regression structure. We believe that the interaction between market orders and limit orders, and the impact of these two types of orders, are crucial to understand the dynamics of the markets, the origin of volatility and the incipient instabilities that can arise when these counteracting forces are not on even keel. The interesting next step would be to analyze in detail these situations, where large liquidity fluctuations arise, and the above `average' model breaks down. On a longer term, a worthwhile project is to construct a coarse-grained, continuous time model from the above microstructural bricks, and justify or reject the slew of models that have been proposed to describe financial time series (L\'evy processes, GARCH, multifractal random walk, etc.).  

\bibliography{limit}

\appendix

\section{The dependence of the spread on the event flow}
\label{app:spread}
In this appendix we show how the framework introduced in the main text can be used to study the 
dynamics of the bid-ask spread. For the spread $S_t$, one can write an exact formula very similar to Eq. (\ref{eq:permanent}):
\begin{equation}
    S_{t+\ell} = S_t + \sum_{t\leq t'< t+\ell} \sum_{\pi} \overline{\Delta}_{\pi, \epsilon_{t'}, t'} I(\pi_{t'}=\pi),
    \label{eq:spread}
\end{equation}
where $\overline{\Delta}_{\pi} = \pm 2\Delta_{\pi}$ with the $+$ sign for $\pi = \MO', \CA'$ and the $-$ sign for $\pi = \LO'$. The other three $\overline{\Delta}$'s are zero, just as the respective $\Delta$'s were. The above equation is accurate because our model includes all the possible events that can change the best quotes, and thus all the possible events that can change the spread. 

However, when formulating a permanent impact model for the spread dynamics in the same spirit as we did for the price, one should bear in
mind that the spread is a mean-reverting quantity that oscillates around a mean value $\ev{S}$. In other words, the average 
value of $S_{t+\ell}$ when $\ell \to \infty$ is equal to $\ev{S}$, independently of the initial value $S_t$. Therefore:
\begin{eqnarray}
\lim_{\ell\rightarrow\infty}\ev{S_{t+\ell}-S_t|S_t} = \lim_{\ell\rightarrow\infty} \sum_{t\leq t'<t+\ell}
\sum_{\pi_2} \ev{\overline{\Delta}_{\pi_2, t', \epsilon_{t'}} I(\pi_{t'}=\pi_2) | S_t } = \ev{S}-S_t.
\label{eq:limitspread}
\end{eqnarray}
Since the right hand side obviously depends on $S_t$, the conditional value 
$\ev{\overline{\Delta}_{\pi_2, t', \epsilon_{t'}} I(\pi_{t'}=\pi_2) | S_t }$ also has to. This means that the event flow
and possibly the gaps are correlated with the spread, and they adjust such that the spread mean reverts. If this were not true, the spread would follow an unbounded random walk. To illustrate this, the spread dependence of realized gaps is shown in Fig. \ref{fig:delta_vs_spread}. 

\addsmallfig{delta_vs_spread}{}{delta_vs_spread}{Realized gaps as a function of the spread (after removing the beginning and the end of the trading days).}

A related study by Ponzi \etal \cite{ponzi.liquidity} based on a selection of stocks from LSE comes to a similar conclusion. They find that both realized gaps and event rates are functions of the spread. In particular limit orders are placed deeper in the spread when the spread is larger. In addition, they show that the rate of transactions decreases with larger spreads, while the rates of cancellations and incoming limit orders increase sharply.

Such an adaptive behavior can be quantified through Eq. \eqref{eq:spread}, for $\ell=1$ this reads
\begin{equation}
    S_{t+1} - S_t = \overline{\Delta}_{\MO', \epsilon_t, t}I(\pi_t=\MO') + \overline{\Delta}_{\CA', \epsilon_t, t}I(\pi_t=\CA') - \left |\overline{\Delta}_{\LO', \epsilon_t, t}\right | I(\pi_t=\LO'),
    \label{eq:spread1}
\end{equation}
where we took the negative absolute value of $\overline{\Delta}_{\LO'}$ to emphasize that it is strictly negative, while all the other quantities in the equation are non-negative. The unconditional expectation value of the left hand side is zero, since $\ev{S_{t+1}}=\ev{S_t}$. Thus on average the spread-altering effect of market orders, cancellations and limit orders balances out. If for example $S_t > \ev{S}$, then the spread mush shrink back to its mean, so the left hand side must be negative. The only way for the right hand side to become negative as well, is if the spread-opening contribution of market orders/cancellations decreases and/or the spread-closing effect of limit orders increases.

This is possible by the variation of both the gaps and the event rates, for our purposes it is enough to introduce a combined description the two. The simplest possible model of such adaptive dynamics is to assume that the conditional distribution of the random variable $\overline{\Delta}_{\pi, \epsilon_t, t}I(\pi_t=\pi)$ given the current spread value $S_t$ can be approximated by that of
$$\overline{\Delta}^\mathrm{R}_{\pi} I(\pi_t=\pi) \left [1 + \frac{\alpha}{\overline{\Delta}^\mathrm{R}_{\pi}} (\ev{S}_\pi-S_{t})\right ],$$ 
where $I(\pi_t=\pi)$ now follows its unconditional distribution, $\ev{S}_\pi$ is the average value of the spread at the time of events of type $\pi$, and $\alpha$ is a constant parameter characterizing the strength of mean-reversion.

Even though the term in the brackets is understood to include contributions from both gap and rate adjustments, technically such a model only amounts to substituting
\begin{equation}
\overline{\Delta}_{\pi, t', \epsilon_{t'}} = \overline{\Delta}^\mathrm{R}_{\pi} + \alpha (\ev{S}_\pi-S_{t'}),
\end{equation}
for the gap dynamics in Eq. \eqref{eq:spread}, and this makes analytical calculations possible. One finds that the modified spread behavior is such that
\begin{eqnarray}
    S_{t+\ell} = S_t (1-\alpha)^\ell + \sum_{t\leq t'< t+\ell} \sum_{\pi} (1-\alpha)^{t+\ell-t'+1} I(\pi_{t'}=\pi) ( \overline{\Delta}^\mathrm{R}_{\pi} + \alpha \ev{S}_\pi),
    \label{eq:spread2}
\end{eqnarray}
from which one deduces the spread response function:
\begin{eqnarray}
      R^S_{\pi_1}(\ell) = \ev{(S_{t+\ell}-S_t)I(\pi_t = \pi_1)}/P(\pi_1) = \nonumber \\ \left [ \ev{S}-\ev{S}_{\pi_1}\right ][1-(1-\alpha)^\ell] + \sum_{t\leq t'< t+\ell} 
    \sum_{\pi_2} (1-\alpha)^{(t+\ell-1)-t'} \overline{\Delta}^\mathrm{R}_{\pi_2} P(\pi_2) \Pi_{\pi_1, \pi_2}(t'-t).
    \label{eq:spreadrf}
\end{eqnarray}
Eq. \eqref{eq:spreadrf} tells us that the dynamics of the spread is related to the autocorrelation of (unsigned) event types, just as the response function was related to the signed event autocorrelation functions (except for the inclusion of $\alpha$ to describe adjustment to the event flow).

To test this model on real data, in order to remove the effect of intraday periodicity and overnight effects, we will neglect the first $30$ and last $40$ minutes of trading days, and in all correlation functions we will only consider times when both events are within the same day. The spread response functions and their approximations \emph{without} allowing for gap fluctuations ($\alpha=0$) are shown in Figs. \ref{fig:redressepsStest023} and \ref{fig:redressepsStest145}. The constant gap approximation works well for large tick stocks, but for small tick stocks only for short times, up to $l\approx 10-30$ events. This is in line with the findings of Sec. \ref{sec:perm} for the response function of price.

\addwidefig{redressepsStest023}{}{redressepsStest023}{Comparison of true and approximated normalized spread response function $R^S_\pi(\ell)/P(\pi)$ for {\it (left)} large tick stocks and {\it (right)} small tick stocks for events that do not change the price. Symbols correspond to the true value, and lines to the approximation. The curves are labeled according to $\pi$ in the legend.}
\addwidefig{redressepsStest145}{}{redressepsStest145}{Comparison of true and approximated normalized spread response function $R^S_\pi(\ell)/P(\pi)$ for {\it (left)} large tick stocks and {\it (right)} small tick stocks for events that change the price. Symbols correspond to the true value, and lines to the approximation. The curves are labeled according to $\pi$ in the legend.}

\begin{figure*}[!tp]
\centerline{\includegraphics[width=220pt,angle=-90,trim=180 0 -50 330]{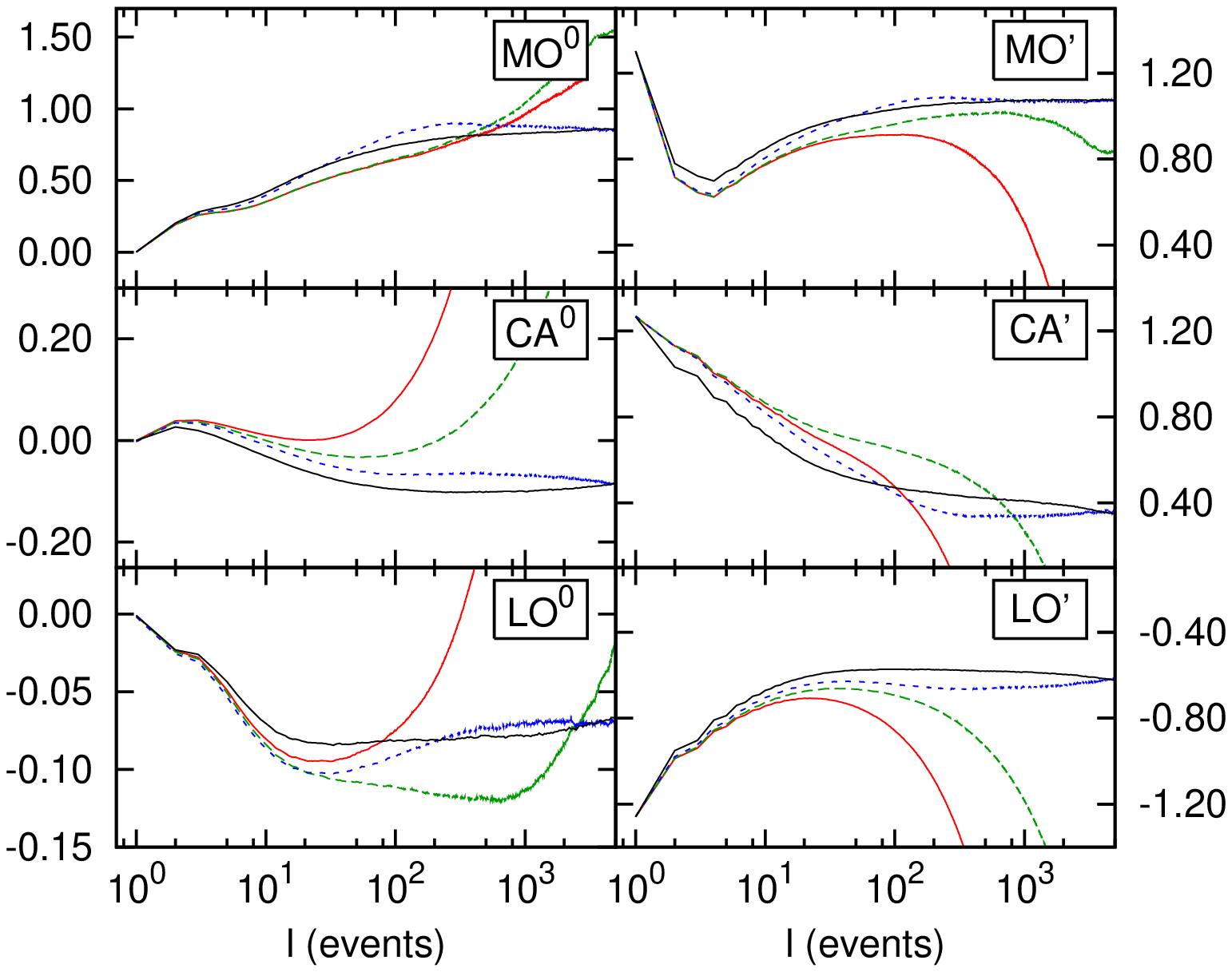}}
\caption{Comparison of the spread response functions of AAPL for four different cases: Eq. \eqref{eq:spreadrf} with $\alpha=0$ (red), Eq. \eqref{eq:spreadrf} with adjusted $\Pi$'s and $\alpha=0$ (green), Eq. \eqref{eq:spreadrf} with adjusted $\Pi$'s and $\alpha=10^{-2}$ (blue), true response functions (black). Price in ticks.}
\label{fig:redressepsSalpha}
\end{figure*}

One finds that the discrepancy for small tick stocks has two origins. First, due to the intraday non-stationarity of the spread the relationship
$$\lim_{\ell\rightarrow\infty} \frac{\overline{\Delta}^\mathrm{R}_{\pi_2} \ev{I(\pi_{t}=\pi_1)I(\pi_{t+\ell}=\pi_2)}}{\ev{\overline{\Delta}_{\pi_2,\epsilon_{t+\ell},t+\ell} I(\pi_{t}=\pi_1)I(\pi_{t+\ell}=\pi_2)}} = 1$$
no longer holds. After excluding the overnight contribution (when $t$ and $t+\ell$ are in different days), the gaps in the denominator are no longer sampled from the first $\ell$ events of the day, where they are systematically larger than at the end of the day. Even after adjusting $\Pi$'s to have the correct asymptotic value, one needs to introduce $\alpha > 0$ to find an approximately correct shape of the spread response functions. The results for one example stock (AAPL) are shown in Fig. \ref{fig:redressepsSalpha}.

Clearly this model is only intended as a first approximation, since it leads to an exponential decay of the spread autocorrelation function in constrast with the long memory found in the data, see Fig. \ref{fig:acspreadexp}. It is possible to give a more complete description of the spread along the lines of Sec. \ref{sec:gap2}, but we will leave this for future research.

\addsmallfig{acspreadexp}{}{acspreadexp}{Autocorrelation function of the spread (after removing the beginning and the end of the trading days). One can see that the exponential decay suggested by our simplified model captures the short-time dynamics, but not the slow decay lasting for thousands of events.}

\section{Plots of various correlation functions}
\label{app:data}

In this appendix we show all the signed and some unsigned correlation functions, signed and unsigned, for small and large tick stocks separately, see Figs.
\ref{fig:acrenorm_largetick_0}-\ref{fig:acpirenorm_smalltick_0}.

\clearpage

\paddtwofigs{acrenorm_largetick_0}{}{acrenorm_largetick_0}{acrenorm_largetick_1}{The normalized, signed event correlation functions $C_{\pi_1, \pi_2}(\ell)$ for large tick stocks, {\it (left)} $\pi_1=\MOO$, {\it (right)} $\pi_1=\MO'$. The curves are labeled by their respective $\pi_2$'s in the legend. The bottom panels show the negative values.}

\paddtwofigs{acrenorm_largetick_2}{}{acrenorm_largetick_2}{acrenorm_largetick_3}{The normalized, signed event correlation functions $C_{\pi_1, \pi_2}(\ell)$ for large tick stocks, {\it (left)} $\pi_1=\CAO$, {\it (right)} $\pi_1=\LOO$. The curves are labeled by their respective $\pi_2$'s in the legend. The bottom panels show the negative values.}

\paddtwofigs{acrenorm_largetick_4}{}{acrenorm_largetick_4}{acrenorm_largetick_5}{The normalized, signed event correlation functions $C_{\pi_1, \pi_2}(\ell)$ for large tick stocks, {\it (left)} $\pi_1=\CA'$, {\it (right)} $\pi_1=\LO'$. The curves are labeled by their respective $\pi_2$'s in the legend. The bottom panels show the negative values.}

\paddtwofigs{acrenorm_smalltick_0}{}{acrenorm_smalltick_0}{acrenorm_smalltick_1}{The normalized, signed event correlation functions $C_{\pi_1, \pi_2}(\ell)$ for small tick stocks, {\it (left)} $\pi_1=\MOO$, {\it (right)} $\pi_1=\MO'$. The curves are labeled by their respective $\pi_2$'s in the legend. The bottom panels show the negative values.}

\paddtwofigs{acrenorm_smalltick_2}{}{acrenorm_smalltick_2}{acrenorm_smalltick_3}{The normalized, signed event correlation functions $C_{\pi_1, \pi_2}(\ell)$ for small tick stocks, {\it (left)} $\pi_1=\CAO$, {\it (right)} $\pi_1=\LOO$. The curves correspond to the six possible values of $\pi_2$'s, see the legend of Fig. \ref{fig:acrenorm_smalltick_0} for details. The bottom panels show the negative values.}

\paddtwofigs{acrenorm_smalltick_4}{}{acrenorm_smalltick_4}{acrenorm_smalltick_5}{The normalized, signed event correlation functions $C_{\pi_1, \pi_2}(\ell)$ for small tick stocks, {\it (left)} $\pi_1=\CA'$, {\it (right)} $\pi_1=\LO'$.  The curves correspond to the six possible values of $\pi_2$'s, see the legend of Fig. \ref{fig:acrenorm_smalltick_0} for details. The bottom panels show the negative values.}

\paddtwofigs{acpirenorm_largetick_0}{}{acpirenorm_largetick_0}{acpirenorm_largetick_1}{The normalized, unsigned event correlation functions $\Pi_{\pi_1, \pi_2}(\ell)$ for large tick stocks, {\it (left)} $\pi_1=\MOO$, {\it (right)} $\pi_1=\MO'$. The curves are labeled by their respective $\pi_2$'s in the legend. The bottom panels show the negative values.}

\paddtwofigs{acpirenorm_smalltick_0}{}{acpirenorm_smalltick_0}{acpirenorm_smalltick_1}{The normalized, unsigned event correlation functions $\Pi_{\pi_1, \pi_2}(\ell)$ for small tick stocks, {\it (left)} $\pi_1=\MOO$, {\it (right)} $\pi_1=\MO'$. The curves correspond to the six possible values of $\pi_2$'s, see the legend of Fig. \ref{fig:acpirenorm_largetick_0} for details. The bottom panels show the negative values.}

\end{document}